\documentclass[a4paper,12pt]{article}
\usepackage{epsfig}
\usepackage{supertabular}
\usepackage{amsmath}
\usepackage{amssymb}
\newcommand{\newc}{\newcommand}
\newc{\eps}{\epsilon}
\newc{\lam}{\lambda}
\newc{\Lam}{\Lambda}
\newc{\ra}{\rightarrow}
\newc{\wtilde}{\widetilde}
\newc{\ie}{{\it i.e.}}
\newc{\rpv}{\not\!\! R_p}
\newc{\lsim}{\stackrel{<}{\sim}}
\newc{\beq}{\begin{equation}}
\newc{\eeq}{\end{equation}}
\newc{\beqn}{\begin{eqnarray}}
\newc{\eeqn}{\end{eqnarray}}
\newc{\PLB}{\emph{Phys.Lett.} {\bf{B}}}
\newc{\NPB}{\emph{Nucl.Phys.} {\bf{B}}}
\newc{\SM}{$S\!M$}
\newc{\ol}{\overline}
\newc{\mcal}{\mathcal}
\newc{\nonum}{\nonumber}
\newc{\bsym}{\boldsymbol}
\newc{\wtil}{\widetilde}
\newc{\lra}{\leftrightarrow}
\newc{\llra}{\longleftrightarrow}
\newc{\dpst}{\displaystyle}
\begin{document}
\setlength{\baselineskip}{.6cm}
\title{\textbf{Supersymmetric Froggatt-Nielsen Models  with Baryon- 
and Lepton-Number Violation}}
\date{}
\author{ \\ Herbi K. Dreiner\footnote{dreiner@th.physik.uni-bonn.de},
$~$ 
            Marc Thormeier\footnote{New address as of March 
10$^{\rm th}$, 
2003:
            \emph{$~$Theoretical Physics Group, $~$Lawrence Berkeley 
National $~~~~~~$Laboratory,  $~$Berkeley,$~\!$CA 94720,  
$~$USA}.$~~~~~~~$E-mail: mhthormeier@lbl.gov}\\
  \\
  \\
\emph{Physikalisches Institut der Universit\"at Bonn,}\\
\emph{Nu\ss allee 12, 53115 Bonn, Germany}} 
\maketitle
\begin{abstract}
$~$\\
We systematically investigate the embedding of $U(1)_X$
Froggatt-Nielsen models in (four-dimensional) local supersymmetry. We
restrict ourselves to models with a single flavon field.  We do not
impose a discrete symmetry by hand, {\it e.g.} $R$-parity,
baryon-parity or lepton-parity.  Thus we determine the order of
magnitude of the baryon- and/or lepton violating coupling constants
through the Froggatt-Nielsen scenario.  We then scrutinize whether
the predicted coupling constants are in accord with weak or GUT scale
constraints. Many models turn out to be incompatible.
\end{abstract}
%
%
%
%
\section{Introduction}
\setcounter{equation}{0}
The problem of the fermionic mass spectrum is unresolved. Within the
Standard Model ($S\!M$), the masses are due to the Yukawa 
interactions
of the Higgs scalar. A theory of fermion masses is thus a theory of
the origin of the Yukawa couplings, a problem not dealt with by the
$S\!M$. This persists when extending the \SM\, to supersymmetry. In
fact the problem becomes worse. Including all terms in the
superpotential which are allowed by the gauge group, there are 45
unknown {\it extra} $R$-parity violating ($\not\!\!R_p$) Yukawa
couplings beyond the 27 of the $S\!M$. A theory addressing the
fermionic mass spectrum should explain all parameters in the
superpotential, {\it i.e.\,} the $\not\!\!R_p$ coupling constants as
well as the bilinear superpotential terms (the
$\mu$-problem) and possible higher-dimensional operators.  There is
extensive experimental data on the conservation of baryon- and
lepton-number and thus bounds on the $\not\!\!R_p$ coupling
constants. It is the purpose of this paper to show how this can be
utilized to constrain the supersymmetric flavour theory, possibly
pointing towards a solution of the fermionic mass problem.

The problem of the fermionic mass spectrum has been extensively
discussed in the literature. In Refs.~[1-16] and Refs.~[17-21], a
certain class of (mostly) supersymmetric Froggatt-Nielsen models
\cite{fn} is presented. The models in Refs.~[1-16] all feature:
\begin{enumerate}
\item A \emph{single}  local $U(1)_X$ symmetry. The gauge charges, 
$X$, 
of the superfields are generation-dependent.
\item This $U(1)_X$  gets broken spontaneously by a single scalar 
flavon field.
\item  The mixed  $U(1)_X$ anomalies are  canceled by the 
Green-Schwarz mechanism \cite{gs}.
\item The fermionic mass spectrum   at the  grand unified 
(GUT)-scale  
satisfies \cite{rrr,n},
\begin{eqnarray}\label{1}
m_e:m_\mu:m_\tau&\sim&\eps^{4~or~5}:\eps^2:1,\\ 
m_\tau:m_b&\sim& 1,\\
m_d:m_s:m_b&\sim&\eps^4:\eps^2:1,\\ 
m_b:m_t&\sim&\eps^{0,1,2~or~3}\;\cot\beta,\label{4}\\ 
m_u:m_c:m_t&\sim&\eps^8:\eps^4:1,\label{5}\\
m_t&\sim&\langle H^\mcal{U}\rangle,\label{NeWW}\\
{\bsym{U^{\!C\!K\!M}}}
&\sim&\left(\begin{array}{lll} 1 & \eps & \eps^3 \\
\eps & 1 & \eps^2 \\
\eps^3 & \eps^2 & 1 \end{array}\right)^{\phantom{M}}~~\mbox{or}~~
\left(\begin{array}{lll}
1 & \eps^2 & \eps^4 \\
\eps^2 & 1 & \eps^2 \\
\eps^4 &\eps^2 & 1 \end{array}\right).\label{6}\\ \nonum
\end{eqnarray}
Here $\tan\beta$ is the ratio of the vacuum expectation values (VEVs)
of the two neutral Higgs scalars, $\tan\beta\equiv\langle H^\mcal
{U}\rangle/\langle H^\mcal{D}\rangle$; ${\bsym{U^{\!C\!K\!M}
}}$ is the Cabibbo-Kobayashi-Maskawa matrix and $\eps\approx0.22$ is
the Wolfenstein parameter, {\it i.e.\,} the sine of the Cabibbo 
angle.
Since the expansion parameter is rather large ($\eps$ and
$\eps^2$ are of the same order of magnitude) the powers of
$\eps$ are only approximate.
\item The models are all in four-dimensional space-time.
\end{enumerate}
The models in Refs.~[1-9] do not contain right-handed neutrinos, the
models in Refs.~[10-16] do. We have tried to present an exhaustive
list of the existing models fulfilling the four points stated above.
One model in Ref.~[16] and the models in Refs.~[17-21] fulfill the
first three points but deviate in their fermionic mass
spectrum.\footnote{We do neither consider a gauged $U(1)$
family-dependent $R$-symmetry as in Ref.~\cite{chamdrei} nor a
discrete gauge symmetry as in Ref.~\cite{babugowa}.}

In this paper, we present a systematic embedding of Froggatt-Nielsen
models in supersymmetry, including a detailed discussion of the
K\"ahlerpotential. We then investigate whether the $U(1)_X$ charge
assignments of the aforesaid models give rise to an acceptable
low-energy phenomenology. We hereby make the following assumptions:
\begin{enumerate}
\item The models are supersymmetric.
\item  Baryon-  \cite{ir} and lepton-parity (the latter one being anomalous) are not imposed by hand,
so {\it a priori} all superpotential terms that are allowed by the
symmetries are assumed to indeed exist; we shall hence include
$\not\!\!R_p$ (for an introduction see {\it e.g.} Refs. 
\cite{rpv1,rpv2,rpv3,rpv4}) and also higher dimensional 
operators.\footnote{In $SO(10)$ grand unified theories, the GUT
symmetry can be broken by a Higgs field in the spinorial {\bf
16}-representation or in the {\bf 126}-representation. If we
exclusively use the latter, then $R$-parity is conserved. If we 
include
the former, then $R$-parity will be violated after $SO(10)$ breaking,
see Refs.~\cite{so101,so102,add3}. 
From the low-energy point of view, 
the choice of 
breaking scheme, however, still appears arbitrary.}
\item  Supersymmetry is local and the breaking of supersymmetry is 
mediated by gravity \cite{gramed1,gramed2,gramed3,gramed4}; hence the
gravitino mass $M_{3/2}$ is of the same order as the masses of the
squarks and sleptons, and the Kim-Nilles \cite{kn} or the
Giudice-Masiero \cite{gm} mechanism solves the $\mu$-problem [this is
consistent with the existence of $U(1)_X$, see Eq.~(\ref{kngm})].
\end{enumerate}
When the $U(1)_X$ is broken ({\it c.f.}\, Section~\ref{FroNie}), the
$R$-parity violating minimal supersymmetric standard model is
generated.  In general, we shall allow for three right-chiral 
neutrino
superfields and we denote the model by $\not\!\!R_p$-$M\!S\!S\!M\!+\!
\ol{N ^i}$. Its most general renormalizable superpotential is 
given by
\begin{eqnarray}\label{superpot}\nonum\\
\mcal{W}~~~=~~~\varepsilon^{ab}~~\delta^{xy}~~{G^{(U)}}_{\!ij}~~Q^i
_{~\!xa}~~{H^{\mcal{U}}}_{\!b}~~\ol{U^j_{~y}}\!\!\!\!\!\!\!\!\!\!
&~&~\nonum\\
+~~\varepsilon^{ab}~~\delta^{xy}~~{G^{(D)}}_{\!ij}~~Q^i_{~\!xa}~~
{H^{\mcal{D}}}_{\!b}~~\ol{D_{~y}^j}\!\!\!\!\!\!\!\!\!\!&~&~
\nonum\\
+~~\varepsilon^{ab}~~{G^{(E)}}_{\!ij}~~L^i_{~\!a}~~
{H^{\mcal{D}}}_{\!b}~~
\ol{E^j}\!\!\!\!\!\!\!\!\!\!&~&~~+~~\varepsilon^{ab}~~
{G^{(N)}}_{\!ij}
~~{L^i}_{\!a}~~{H^{\mcal{U}}}_{\!b}~~\ol{N^j}\nonum\\
~~~~~~~~~~~~~~~~+~~\varepsilon^{ab}~~\mu~~ {H^{\mcal{D}}}_{\!a}~~
{H^{\mcal{U}}}
_{\!b}&~&~~+~~\Gamma_{ij}~~\ol{N^i}~~\ol{N^j}\nonum\\
\nonum\\
+~~\frac{1}{2}~~\varepsilon^{ab}~~\Lam_{ijk}~~{L^i}_{\!a}~~
{L^j}_{\!b}~~
\ol{E^k}\!\!\!\!\!\!\!\!\!\!&~&\nonum\\
+~~\varepsilon^{ab}~~\delta^{xy}~~{\Lam^{\prime}}_{ijk}~~
{Q^i}_{\!xa}~~
{L^j}_{\!b}~~
\ol{D^k_{~y}}\!\!\!\!\!\!\!\!\!\!&~&~~+~~\Xi_i~~\ol{N^i}   
\nonum\\
+~~\frac{1}{2}~~\varepsilon^{xyz}~~{\Lam^{\prime\prime}}_{ijk}~~
\ol{U^i_{~x}}~
\ol{D^j_{~y}}~~\ol{D^k_{~z}}\!\!\!\!\!\!\!\!\!\!&~&~~+~~
\varepsilon^{ab}~~
{\Upsilon}_i~~ \ol{N^i}~~{H^{\mcal{D}}}_{\!a} ~{H^{\mcal{U}}}_
{\!b} 
\nonum\\
+~~\varepsilon^{ab}~~K_i~~{L^i}_{\!a}~~{H^{\mcal{U}}}_{\!b}\!\!
\!\!\!\!\!\!
\!\!&~&~~+~~{\Lam^{\prime\prime\prime}}_{ijk}~~\ol{N^i}~~
\ol{N^j}~~
\ol{N^k}.\\ \nonum
\end{eqnarray}
We use the standard notation, which is explained in the footnote
below.\footnote{$H,Q,L$ represent the left-chiral $SU(2)_W$-doublet
superfields of the Higgses, the quarks and leptons; $U,D,E,N$
represent the right-chiral superfields of the $u$-type quarks, 
$d$-type
quarks, electron-type and neutrino-type leptons, respectively; an
overbar denotes charge conjugation; $a,b,c$ and $x,y,z$ are
$SU(2)_{W}$- and $SU(3)_{C}$-indices, respectively, $~i,j,k,l$ are
generational indices, summation over all repeated indices is implied;
$\delta^{xy}$ is the Kronecker symbol, $\varepsilon^{...}$ symbolizes
any tensor that is totally antisymmetric with respect to the exchange
of any two indices, with $\varepsilon^{12...}=1$.  All other symbols
are coupling constants, {\it a priori} they are arrays of arbitrary
complex numbers (except $\Lam_{ijk}/{\Lam^{\prime\prime}}_{ijk}$
being antisymmetric with respect to the exchange of the first 
two/last
two indices, $\Gamma_{ij}$ and ${\Lam^{\prime\prime\prime}}_{ijk}$
being totally symmetric with respect to any exchange of indices).} 
The upper two blocks in
Eq.~(\ref{superpot}) are $R_p$ conserving, the lower blocks violate
$R_p$; the left blocks do not contain right-handed neutrinos, the 
right
blocks do (hence they are absent in the models of Refs.~[1-9]).  
Thus the
superpotential of the $M\!S\!S\!M$ is contained in the upper left
block.  We also consider non-renormalizable 
$F$-term operators of
dimensionality five; having dropped all gauge and flavour indices 
they
read
\begin{eqnarray}\label{nonrenorm}
LH^\mcal{U}LH^\mcal{U},~~QQ\ol{U}\ol{D},~~ QL\ol{U}
\ol{E},
 &~~ &  \ol{N}LL\ol{E},~~ \ol{N}QL
\ol{D},
~~\ol{N}\ol{U}\ol{D}\ol{D},\nonum\\
QQQL,~~H^\mcal{D}H^\mcal{U}H^\mcal{D}H^\mcal{U},~~ 
\ol{UUDE},
       &~~ &  \ol{N}
\ol{N}
H^\mcal{D}H^\mcal{U}, ~~\ol{NNNN},\nonum \\
                                      &~~ &  \nonum \\          
LH^\mcal{U}H^\mcal{D}H^\mcal{U},~~QH^\mcal{D}\ol{U}
\ol{E},
~~QQQH^\mcal{D},           &~~&    \ol{N}QH^\mcal{U}
\ol{U},~~ 
\ol{N}QH^\mcal{D}\ol{D}, ~~\ol{N}LH^\mcal{D}\ol{E},
 \nonum\\ & &\ol{NN}LH^\mcal{U}\,,
\end{eqnarray}
grouped according to the same scheme as in Eq.~(\ref{superpot}). 
Note that there are operators which conserve $R_p$ but violate
baryon- and lepton-parity unlike the case with renormalizable 
operators.
We
shall focus on those operators that can lead to rapid proton decay,
\begin{equation}\label{QQQL}
\frac{\bsym{\Psi}}{M_{\!s}}QQQL,~
\frac{\bsym{\Psi^\prime}}{M_{\!s}}QQQH^\mathcal{D};
\end{equation}
$M_{\!s}$ is the string scale ($\approx10^{18}$GeV),
 $\bsym{\Psi},\bsym{\Psi^\prime}$ are arrays of  dimensionless 
numbers. We do not
consider the operators $\ol{UUDE}$, because they contribute to proton
decay only via the unknown quark-squark mixing matrix for the
right-handed up-quarks, see for example Refs.~\cite{hk,hpn,bhn}.

This text is structured as follows. In Section~\ref{FroNie}, we 
review
the idea of Froggatt and Nielsen within supersymmetry and consider
some of its implications. In Section~\ref{constr}, we discuss the
phenomenological constraints on the charges of the $U(1)_X$
symmetry. Then we parameterize and list the most general set of
$X$-charges, consistent with anomaly cancellation and
Eqs.~(\ref{1}-\ref{6}). In Section~\ref{procedure}, we explain the
procedure with which we get from the string scale where the model is
naturally formulated down to the weak scale, as well as which
approximations we utilize. In Section~5 together with
Appendix~\ref{catalogue}, we present and discuss our results. In
Section 6 we offer a short summary.  In Appendix~\ref{catalogue} the
models of Refs.~[1-16] are discussed and listed, as well as the
$X$-charge assignments of the other models. In
Appendix~\ref{illustration} we demonstrate the validity of the mass
matrices we use in this paper.

\section{\label{FroNie}The Scenario of Froggatt and Nielsen} 
In the models we consider, the superpotential and the
K\"ahlerpotential are expected to originate from a ten-dimensional
heterotic superstring, see {\it e.g.}  Ref.~\cite{nillestalk}: After
compactification to four dimensions and below the string cut-off $M_
{\!s}$ such a theory may exhibit at most\footnote{If there IS more
than one anomalous  $U(1)$, one can re-express the charges such that 
only one    $U(1)$ is anomalous. For an explicit example see e.g. 
Ref.~\cite{faraggithormeier} and references therein.} one anomalous 
local 
$U(1)_X$
(see {\it e.g.} Refs.~\cite{kona,far} and references therein). We
assume this anomalous local $U(1)_X$ indeed exists and has
generation-dependent charges. For example, $X_{\!Q^i},$ $i=1, 2,3$,
denote the $U(1)_X$ charges of the quark doublet superfields $Q^i$ 
and
correspondingly for the other superfields.  In addition to $U(1)_X$,
we assume an $X$-charged $S\!M$-singlet left-chiral superfield $A$, 
the
so-called flavon superfield. For simplicity we choose the convention
that $X_A=-1$.

Close to the string scale where $U(1)_X$ is unbroken, we have to
replace the superpotential by a $U(1)_X$ gauge invariant extension,
{\it e.g.} instead of ${G^{(U)}}_{\!ij}\,Q^i\,H^{\mcal{U}}\,\ol{U^j}$
we have the non-renormalizable operators
\begin{eqnarray}\label{2}
{g^{(U)}}_{\!ij}\,Q^i\,H^{\mcal{U}}\,\ol{U^j}\,
\Bigg(\frac{A}{M_{\!s}}\Bigg)^{\!\!{X_{Q^i}+X_{H^{\mcal{U}}}+
X_{\ol{U^j}}}}\;\wtil{\Theta}\Big[X_{Q^i}+X_{H^{\mcal{U}}}+
X_{\ol{U^j}}\Big].
\end{eqnarray}
The complex couplings ${g^{(U)}}_{\!ij}$ are assumed to be of
$\mcal{O}\!(1)$, {\it i.e.}
\begin{equation}\label{Range}
\frac{1}{\sqrt{10}}\lesssim \big|{g^{(U)}}_{\!ij}\big| \lesssim 
\sqrt{10}\,.
\end{equation}
Furthermore $\wtil{\Theta}[x]$ is defined as
\begin{equation}
\wtil{\Theta}[x]\equiv\left\{
\begin{array}{cl} 
1 & \;\mbox{for $x$ being a non-negative integer\,,}  \\
& \\ 
0  & \;\mbox{else (supersymmetric zero)\,.} \end{array}\right. 
\end{equation}
$\wtil{\Theta}[...]$ appears in Eq.~(\ref{2}) due to: {\it (i)}
Holomorphicity of the superpotential: negative exponents of the
superfield $A$ are not allowed. The resulting texture is denoted a
{\it  supersymmetric zero} \cite{lns1}. {\it (ii)} 
Fractional exponents are also forbidden, somewhat imprecisely the
corresponding zeros shall be referred to as supersymmetric zeros as
well. If an exponent of $\frac{A}{M_{\!s}}$ is fractional and/or
negative it shall be referred to as a {\it naive
exponent}.\footnote{However, as stated in Ref.~\cite{br} one could
``envisage fractional powers of the [super]field [$A$], stemming from
non-perturbative effects''.} The superpotential terms $\mu
H^\mcal{D}H^\mcal{U}$ and $K_iL^iH^\mcal{U}$ are treated in
exactly the same manner:
\begin{eqnarray}\label{gm}
M^{(\mu)}~~H^\mcal{D}~H^\mcal{U}~~\Bigg(\frac{A}{M_{\!s}}
\Bigg)^{\!\!{X_{H^\mcal{D}}+X_{H^{\mcal{U}}}}}~~
\wtil{\Theta}\Big[X_{H^\mcal{D}}+X_{H^{\mcal{U}}}\Big],
\nonum\\
M^{(\mu)}~~\kappa_i~~L^i~H^\mcal{U}~~\Bigg(\frac{A}{M_{\!s}}
\Bigg)^{\!\!{X_{L^i}+X_{H^{\mcal{U}}}}}~~\wtil{\Theta}
\Big[X_{L^i}+X_{H^{\mcal{U}}}\Big].
\end{eqnarray}
$M^{(\mu)}$ is a dimensionful parameter, the $\kappa_i$ are
dimensionless and of $\mcal{O}\!(1)$. Note that the same mass scale 
is
assumed in both expressions above. Likewise for the superpotential
terms $\Xi_i\,\ol{N^i}$ and $\Gamma_{ij}\,\ol{N^i}~\ol{N^j}$. We 
again
assume a common mass scale, in general different from $M^{(\mu)}$:
\begin{eqnarray}\label{gm2}
&&\Big(M^{(\ol{N})}\Big)^2\,\xi_i\,\ol{N^i}\;
\Bigg(\frac{A}{M_{\!s}}\Bigg)^{\!\!{X_{\ol{N^i}}}}\;
\wtil{\Theta}\Big[ X_{\ol{N^i}}  \Big],\nonum\\
&&\;M^{(\ol{N})}\;\gamma_{ij}\;\ol{N^i}\;
\ol{N^j}\;\Bigg(\frac{A}{M_{\!s}}\Bigg)^{\!\!{X_{\ol{N^i}}
+X_{\ol{N^j}}}}\;\wtil{\Theta}\Big[X_{\ol{N^i}} 
+X_{\ol{N^j}} \Big].
\end{eqnarray}
$M^{(\ol{N})}$ is a dimensionful parameter presumably of $\mcal{O}
(M_{\rm GUT}$) or $\mcal{O}(M_{\!s}$) , the $\xi_i,\gamma_{ij}$ are
dimensionless and of $\mcal{O}\!(1)$. If one is not willing to
introduce yet another mass scale $M^{(\ol{N})}$ in addition to
$M_{3/2}$ and $M_{\!s}$ (and possibly $M^{(\mu)}$), one can set
$M^{(\ol{N})}=M_{\!s}$ and sufficiently suppress the righthanded
neutrino masses by powers of $\epsilon$. We will explicitly deal with
this point in Ref.~\cite{dmt}. For a demonstration see the remarks
about Ref.~[10] in Appendix~A, in particular Eq.~(\ref{zem}).

Just like the superpotential, the K\"ahlerpotential also has to be
replaced, {\it e.g.} instead of $~{H^{(Q)}}_{\!ij}~~
\ol{Q^i}~Q^j$ we have
\begin{eqnarray}\label{hq}
{h^{(Q)}}_{\!ij}\,\ol{Q^{i}}~Q^j\times\Bigg\{\Bigg(\frac{A}{M_{\!s}}
\Bigg)^{X_{Q^j}-X_{Q^i}}\,
\wtil{\Theta}\Big[X_{Q^j}-X_{Q^i}\Big]+\Bigg(\frac{
\ol{A}}{M_{\!s}}\Bigg)^{X_{Q^i}-X_{Q^j}}\,
\wtil{\Theta}\Big[{X_{Q^i}-X_{Q^j}}
\Big]\Bigg\}.\;\;~~~~~
\end{eqnarray}
The complex couplings ${h^{(Q)}}_{\!ij}$ are of $\mcal{O}\!(1)$ and
form a positive-definite Hermitian matrix. For simplicity the gauge
connection matrix is not written out as it is of no importance
here. Note that these terms are not of the canonical form of the
kinetic energy, which is important later on ({\it c.f.} Step 2 in
Section 4).

In addition to the expression above, we also obtain K\"ahlerpotential
terms of the form
\begin{eqnarray}\label{kngm}
\!\!\!&&C_{\!\ol{N^i}}\,\ol{Z}\,\ol{N^i}\Bigg
\{\Bigg(\frac{\ol{A}}{M_{\!s}}\Bigg)^{{  -X_{
\ol{N^i}}}}\,\wtil{\Theta}\Big[-X_{\ol{N^i}}\Big]
+ \Bigg(\frac{{A}}{M_{\!s}}\Bigg)^{{  X_{\ol{N^i}}}}
\wtil{\Theta}\Big[X_{\ol{N^i}}\Big]  ~\Bigg\} \, , \nonum\\
\!\!\!&&C_{\!H^\mcal{D}\!H^\mcal{U}}\,\frac{\ol{Z}}
{M_{\!s}} H^\mcal{D}\,H^\mcal{U}\Bigg\{\Bigg(
\frac{\ol{A}}{M_{\!s}}\Bigg)^{{\!\!-X_{H^\mcal{D}}-
X_{H^{\mcal{U}}}}}\,\wtil{\Theta}\Big[-X_{H^\mcal{D}}
-X_{H^{\mcal{U}}}\Big]~+~...~\Bigg\}\,, \\
\!\!\!&&C_{\!L^i\!H^\mcal{U}}\,\frac{\ol{Z}}{M_{\!s}} 
L^i~H^\mcal{U}\Bigg\{\Bigg(\frac{\ol{A}}{M_{\!s}}
\Bigg)^{{\!\!-X_{L^i}-X_{H^{\mcal{U}}}}}\wtil{\Theta}
\Big[-X_{L^i}-X_{H^{\mcal{U}}}\Big]~+~...~\Bigg\}\,,\nonum\\
\!\!\!&&C_{\!Q^i\!H^\mcal{U}\!\ol{U^j}}\,\frac{
\ol{Z}}{{M_{\!s}}^2}\,{Q^i~H^\mcal{U}~\ol{U^j}}
\Bigg\{\Bigg(\frac{\ol{A}}{M_{\!s}}\Bigg)^{{\!\!-X_{Q^i}-
X_{H^{\mcal{U}}}-X_{\ol{U^j}}}}\,\wtil{\Theta}
\Big[-X_{Q^i}-X_{H^{\mcal{U}}}-X_{\ol{U^j}}\Big]~+
~...~\Bigg\},\nonum
\end{eqnarray}  
together with the corresponding complex conjugate terms. The
$C_{\ldots}$ symbolize the dimensionless $\mcal{O}\!(1)$ coupling
constants. $Z$ is an $X$-uncharged left-chiral superfield of the 
hidden
sector, its $F$-term breaks supersymmetry.

Just as with the $U(1)_{X}$ itself, the {\it breaking} of the
$U(1)_X$ is also embedded in string theory. This can be traced back 
to
the necessity to get rid of the mixed chiral anomalies of $U(1)_X$
with the $S\!M$ gauge group.  They are canceled by the
four-dimensional field theory remnant of the
Green-Schwarz mechanism \cite{gs}, relying on two assumptions:
\begin{enumerate}
\item   The left-chiral scalar complex  dilaton superfield $S$
 shifts under a $U(1)_X$ gauge transformation, namely
\begin{eqnarray}\label{dilsh}
S~\rightarrow~ S~+~\frac{i}{2}~\delta_{\rm GS}~\Lambda_X,
\end{eqnarray}
while the other superfields transform the usual way, {\it e.g.} $H^
\mathcal{U}\rightarrow\exp({iX_{H^\mathcal{U}}\Lambda_X})
H^\mathcal{U}$,
$V_X\rightarrow V_X+\frac{i}{2}(\Lambda_X-\ol{\Lambda_X})$. Here, $V_
X$ is the $U(1)_X$ gauge vector superfield, $\Lam_X$ is the 
left-chiral superfield parameterizing the extended $U(1)_X$ gauge
transformations and $\delta_{\rm GS}$ is a real number.
\item   The coefficient of the gravity-gravity-$U(1)_X$ anomaly, 
$\mcal{A}_{\rm GGX}$, as well as the coefficients for the mixed
gauge anomalies
$SU(3)_{C}\mbox{-}SU(3)_{C}\mbox{-}U(1)_{X}$,
$SU(2)_{W}\mbox{-}SU(2)_{W}\mbox{-} U(1)_{X}$,
$U(1)_{Y}\mbox{-}U(1)_{Y}\mbox{-}U(1)_{X}$ and
$U(1)_{X}\mbox{-}U(1)_{X}\mbox{-}U(1)_{X}$ should fulfill, see 
{\it e.g.}
Ref.~\cite{maekawa},
\begin{eqnarray}\label{cancel}
\frac{\mcal{A}_{\rm CCX}}{k_{\!C}}\stackrel{!}{=}
\frac{\mcal{A}_{\rm WWX}}{k_{\!W}}\stackrel{!}{=}
\frac{\mcal{A}_{\rm YYX}}{k_{\!Y}}\stackrel{!}{=}
\frac{\mcal{A}_{\rm XXX}}{3~k_{\!X}}\stackrel{!}{=}
\frac{\mcal{A}_{\rm GGX}}{12}\stackrel{!}{=}~2\pi^2
~\delta_{\rm GS}\,.
\end{eqnarray}  
The $k_{\ldots}$ are the affine or Ka{\v{c}}-Moody levels of the
corresponding symmetry. The most easily constructed string models and
thus so far most string models to date have\footnote{The $k_{...}$
 for
non-Abelian groups have to be positive integers. For gauge theories
arising from string theories $k_{...}~{g_{...}}^2=2{g_{\!s}}^2$;
furthermore heterotic string theory always --- no matter whether one
has a GUT or not --- gives rise to unification of the coupling
constants, see Ref.~\cite{94}. Hence at high energies $g_{Y}=\sqrt
{3/5}\cdot g_{W}$, and thus $k_{Y}=5/3~k_{W}$.}
\begin{equation}\label{kac}
k_{\!C}=k_{\!W},~~~k_{\!Y}=5/3~k_{\!W}\,.
\end{equation}
The anomaly coefficients are given in terms of the 
$X$-charges as
\begin{eqnarray}\label{ccx}
\mcal{A}_{\rm CCX}=\sum_{i=1}^3\Big(2~X_{Q^i}+
X_{\ol{U^i}}+X_{\ol{D^i}}\Big),
\end{eqnarray}
\begin{eqnarray}\label{wwx}
\mcal{A}_{\rm WWX}=X_{H^\mcal{U}}+X_{H^\mcal{D}}
+\sum_{i=1}^3\Big(3~X_{Q^i}+X_{L^i}\Big),
\end{eqnarray}
\begin{eqnarray}\label{yyx}
\mcal{A}_{\rm YYX}=X_{H^\mcal{U}}+
X_{H^\mcal{D}}+\frac{1}{3}\sum_{i=1}^3\Big(X_{Q^i}+
8~X_{\ol{U^i}}+2~X_{\ol{D^i}}+3~X_{L^i}+6
X_{\ol{E^i}}\Big).~~~~
\end{eqnarray}
Due to the possible existence of $X$-charged $S\!M$-singlets other
than $A$ and $\ol{N^i}$, the anomaly coefficients $\mcal{A}_{\rm XXX}
={\rm Tr}X^3$ and $\mcal{A}_{\rm GGX}={\rm Tr}X$ are not helpful to
find constraints on the $X$-charges of the $S\!M$, so they are not
listed.
\end{enumerate}
These two assumptions together ensure that the shift of the dilaton
eliminates the anomalies of the $X$-charged current, because the
dilaton couples universally to the topological terms $\eps^{\mu\nu
\alpha\beta}~k_{...}  ~F_{\mu\nu}^{...}~F_{\alpha\beta}^{...}$ of 
the various gauge groups. For more details see {\it e.g.} 
Ref.~\cite{maekawa,ra}.

To be entirely anomaly-free, the mixed anomaly $U(1)_{\!Y}\mbox{-}U(1
)_ X\mbox{-}U(1)_X$ must vanish on its own,\footnote{Note that if the
$S\!M$ gauge group derives from an $SU(5)$-GUT, {\it i.e.} 
$X_{Q^i}=X_
{\ol{ U^i}}=X_{\ol{E^i}}$, $X_{L^i}=X_{\ol{D^i}}$, and furthermore
$X_{H^\mcal{D}}=-X_{H^\mcal{U}}$, one automatically has $\mcal{A}_{
\rm CCX}=\mcal{A}_{\rm WWX}=\frac{3}{5}\mcal{A}_{\rm YYX}$ and $\mcal
{A}_{\rm YXX}=0$.}
\begin{eqnarray}\label{yxx}      
\mcal{A}_{\rm YXX}={X_{H^\mcal{U}}}^2-{X_{H^\mcal{D}}}^2+
\sum_{i=1}^3\Big({X_{Q^i}}^2-2~{X_{\ol{U^i}}}^2+
{X_{\ol{D^i}}}^2-{X_{L^i}}^2+{X_{\ol{E^i}}}^2\Big)
\stackrel{!}{=}0\,.
\end{eqnarray}
It is crucial that Assumption~1 makes it necessary to have a modified
K\"ahlerpotential for $S$, in order to have $U(1)_X$ gauge 
invariance:
instead of being a function of $~S+\ol{S}~$ it has to be a 
function of
\begin{equation} 
S~+~\ol{S}~-~\delta_{GS}~{V}_X\,.
\end{equation}
This ``new''
K\"ahlerpotential inevitably leads to string radiative corrections
generating a finite $U(1)_X$ 
Fayet-Iliopoulos term  
\cite{dsw0,dsw,ads,ads2}, its coefficient reads
\begin{eqnarray}\label{elf}
{{\xi^{F\!I}}_X}~=
\frac{{g_{\!s}}^2}{192~\pi^2}~{M_{\!s}}^2~\mcal{A}_{\rm GGX}\,.
\end{eqnarray}
${{\xi^{F\!I}}_X}^{\rm tree~ level}$ has to vanish in local
supersymmetry, see Ref.~\cite{barbieri}.  Despite ${{\xi^{F\!I}}_X}
\neq0$ we demand that {\it (i)} $ SU(3)_C\!\times\!SU(2)_W\! \times
\! U(1)_Y$ and {\it (ii)} supersymmetry both remain unbroken at this 
scale. {\it (i)} is the case if the scalar components of all $S\!M
$-superfields have vanishing VEVs. Taking this into account, {\it
(ii)} is given if at least one of the $X$-charges of the remaining
superfields whose scalar component gets a VEV has the opposite sign
 of
${{\xi^{F\!I}}_X}$. This one superfield is the flavon superfield $A$,
its scalar component acquiring the VEV (thereby breaking $U(1)_X$)
denoted by $\upsilon\propto\sqrt{{\xi^{F\!I}}_X}$, see also
footnote~$^{\ref{six10}}$.  Thus the string radiative correction is
essential for the breaking of $U(1)_X$, hence occurring near the
string scale. This is why $\upsilon/M_{\!s}\propto\sqrt{\mathcal{A}
_{\rm{GGX}}/192\pi^2}$ may not be a {\it very} small number.\footnote
{\label{fu9}As can be seen from Eq.~(\ref{elf}), if $U(1)_X$ is
non-anomalous (thus $\mcal {A}_{\rm GGX}=0$) the string radiative
correction vanishes; furthermore if in this case one chooses ${{\xi^{F
\!I}}_X}^{\rm tree~ level}$ appropriately one can break $U(1)_X$ at a much 
lower scale; such a model is sketched in Ref.~\cite{mnr}.} On the
other hand it must be ensured that $\upsilon/M_{\!s}$ is small enough
so that we can identify
\begin{eqnarray}
\frac{\upsilon}{M_{\!s}}\equiv\eps=0.22\,.
\end{eqnarray}
Thus the $X$-charges should be moderately valued, otherwise $\mcal
{A}_{\rm GGX}$ would be so large (assuming that there is no 
fine-tuned
cancellation among the $X$-charges) that it cannot be adequately
suppressed by $192\pi^2$.\footnote{As was shown in 
Ref.~\cite{9612442}
(see also Ref.~\cite{buk}), with Eqs.~(\ref{cancel}, \ref{ccx(xz)})
and using $k_{...}~{g_{...}}^2= {g_s}^2$, $\eps$ is indeed of the 
correct
order of magnitude. See also Ref.~\cite{dmt}.}

After the breaking of $U(1)_X$ we get from Eqs.~(\ref{superpot})
and (\ref{2}) that
\begin{eqnarray}\label{gu}
{G^{(U)}}_{\!ij}\,=\,{g^{(U)}}_{\!ij}~~\eps^{X_{Q^i}+
X_{H^{\mcal{U}}}+X_{\ol{U^j}}}~~\wtil{\Theta}
\Big[X_{Q^i}+X_{H^{\mcal{U}}}+X_{\ol{U^j}}\Big]\,.
\end{eqnarray}
Similarly, Eq.~(\ref{hq}) leads to 
\begin{eqnarray}\label{184}
{H^{(Q)}}_{\!ij}\,=\,{h^{(Q)}}_{\!ij}~~\eps^{|X_{Q^i}-X_{Q^j}|}
~~\wtil{\Theta}\Big[ |X_{Q^i}-X_{Q^j}|   \Big]\,,
\end{eqnarray}  
and likewise for the other coupling constants. Depending on the
$U(1)_X$ charges, {\it e.g.} the sum $X_{Q^i}+X_{H^{\mcal{U}}}+
X_{{\ol
{U}}^j}\,$, we get generation-dependent exponents of $\eps$ and
thus generation-dependent suppressions of the ${G^{(U)}}_{\!ij}\,$.
We thus have generation-dependent hierarchical coupling constants. It
should be emphasized that the idea of Froggatt and Nielsen does not
reduce the number of parameters, but only explains their hierarchy.
As $\eps$ is rather large, instead of the constraint in
Eq.~(\ref{Range}) we shall restrict ourselves to
\begin{equation}\label{Rrrange}
\sqrt{\eps~}~~\lesssim~~ \big|{g^{(U)}}_{\!ij}\big|~~ \lesssim~~ 
\frac{1}{\sqrt{\eps~}}\;.
\end{equation}

Due to the Giudice-Masiero mechanism, K\"ahlerpotential terms as in
Eq.~(\ref{kngm}) also contribute
to the effective superpotential,\footnote{Supersymmetry is broken by
the VEV of $F_Z$, where $F_Z$ is the auxiliary field of $Z$. Thus one
gets $\big\langle\ol{F_Z}\big\rangle~\delta^2(\bar{\theta})=M_{3/2}~
M_{\!s}~\delta^2(\bar{\theta})$. The $\delta$-function turns the
original K\"ahlerpotential term into a superpotential term.} giving
\begin{eqnarray}\label{gfgfgfgf}
M_{\!s}~M_{3/2}&&C_{\!\ol{N^i}}~~\eps^{{|X_{
\ol{N^i}}}|}~~\wtil{\Theta}\Big[|X_{\ol{N^i}}|\Big]~~
\ol{N^i}\,  ,\\
M_{3/2}&&C_{\!H^\mcal{D}\!H^\mcal{U}}~~\eps^{{|
X_{H^\mcal{D}}+X_{H^{\mcal{U}}}}|}~~\wtil{\Theta}
\Big[|X_{H^\mcal{D}}+X_{H^{\mcal{U}}}|\Big]~~H^\mcal{D}~
H^\mcal{U}\,,\nonum\\
M_{3/2}&&C_{\!L^i\!H^\mcal{U}}~~\eps^{{|X_{L^i}+X_{H^{
\mcal{U}}}}|}~~\wtil{\Theta}\Big[|X_{L^i}+X_{H^{
\mcal{U}}}|\Big]~~L^i~H^\mcal{U}\,,\nonum\\
M_{3/2}&&C_{\!\ol{N^i}\ol{N^j}}~~\eps^{{|
X_{\ol{N^i}}+X_{\ol{N^j}}}|}~~\wtil{\Theta}
\Big[{|X_{\ol{N^i}}+X_{\ol{N^j}}}|   \Big]~~
\ol{N^i}~\ol{N^j}\,  ,\nonum\\
\frac{M_{3/2}}{M_{\!s}}&&C_{\!Q^i\!H^\mcal{U}\!\ol{U^j}}
~~\eps^{{|X_{Q^i}+X_{H^{\mcal{U}}}+X_{\ol{U^j}}}|}~~
\wtil{\Theta}\Big[|X_{Q^i}+X_{H^{\mcal{U}}}+X_{\ol{U^j}}|
\Big]~~ Q^i~H^\mcal{U}~\ol{U^j}\,.\nonum
\end{eqnarray} 
It is safe to ignore the contributions to the trilinear 
superpotential
terms, as they are strongly suppressed by at least a factor 
of $\mcal{
O }(10^{-15})$,\footnote{See however Refs. 
\cite{murayamaundco1,murayamaundco2}: They use
the Giudice-Masiero mechanism to get small neutrino masses.} so that
coupling constants as in Eq.~(\ref{gu}) remain unchanged. This is not
the case for the linear and bilinear terms: using the expressions
above and Eqs.~(\ref{gm}, \ref{gm2}) we obtain
\begin{eqnarray}\label{xigm}
\Xi_i&=&\Big(M^{(\ol{N})}\Big)^2\,\xi_i\,\eps^{X_{
\ol{N^i}}}\,\wtil{\Theta}\Big[X_{\ol{N^i}}\Big]
+M_{3/2}\,M_{\!s}\,C_{\ol{N^i}}\,
\eps^{|X_{\ol{N^i}}|}\,\wtil{\Theta}\Big[|
X_{\ol{N^i}}|\Big]\,,
\end{eqnarray}  
\begin{eqnarray}
\Gamma_{ij}&=&M^{(\ol{N})}~~\gamma_{ij}~~\eps^{
X_{\ol{N^i}}+X_{\ol{N^j}}}~~\wtil{\Theta}
\Big[X_{\ol{N^i}}+X_{\ol{N^j}}\Big]\nonum\\
&~&~~+~~M_{3/2}~~C_{\ol{N^i}\ol{N^i}}~~\eps^{
|X_{\ol{N^i}}+X_{\ol{N^j}}|}~~\wtil{\Theta}
\Big[|X_{\ol{N^i}}+X_{\ol{N^j}}|\Big],~
\end{eqnarray}
\begin{eqnarray}\label{mue}
\mu&=&M^{(\mu)}~~\eps^{X_{H^\mcal{D}}+{X_{H^\mcal{U}}}}
~~\wtil{\Theta}\Big[X_{H^\mcal{D}}+{X_{H^\mcal{U}}}
\Big]\nonum\\
&~&~~+~~M_{3/2}~~C_{\!H^\mcal{D}\!H^\mcal{U}}~~\eps^{
|X_{H^\mcal{D}}+X_{H^\mcal{U}}|}~~\wtil{\Theta}\Big[|
X_{H^\mcal{D}}+X_{H^\mcal{U}}|\Big],\\\nonum\\
\label{kbeforefn}
K_i&=&M^{(\mu)}\,\eps^{X_{L^i}+X_{H^\mcal{U}}}\,
\wtil{\Theta}\Big[X_{L^i}+X_{H^\mcal{U}}\Big] \nonum \\
&&+M_{3/2}\;C_{\!L^i\!H^\mcal{U}}\,\eps^{|X_{L^i}+
X_{H^\mcal{U}}|}\,\wtil{\Theta}\Big[|X_{L^i}+{X_{H^
\mcal{U}}}|\Big].
\end{eqnarray}

\medskip

To summarize this section, the supergravity-embedded models which we
consider are determined by the superpotential given in
Eqs.~(\ref{superpot}) and (\ref{QQQL}) together with the
K\"ahlerpotential.  The two potentials are generated close to the
string scale by the scalar component of the flavon superfield
acquiring a VEV. After $U(1)_X$ symmetry breaking, non-renormalizable
operators generate renormalizable operators which are suppressed by
powers of $\eps$ times an unknown $\mcal{O}\!(1)$ coupling
constant. The power of $\eps$ is fixed by the generation-dependent
quantum numbers of the superfields under the $U(1)_X$ family
symmetry. One also has to take into account that via the
Giudice-Masiero mechanism \cite{gm} terms of the K\"ahlerpotential 
can
contribute to the superpotential after $U(1)_X$ symmetry
breaking. Numerically the effects are only relevant for the linear 
and
bilinear terms.
\section{\label{constr}Phenomenological Aspects}
  
We would next like to derive a conclusion from the previous section
(see also Ref.~\cite{iba}). Consider Eq.~(\ref{cancel}), it leads to
\begin{eqnarray}\label{ibanez}
\frac{1}{2}~\Big({\mcal{A}_{\rm YYX}}+ 
\mcal{A}_{\rm WWX}-\frac{8}{3}~\mcal{A}_{\rm CCX}\Big)&=& 
 \frac{\mcal{A}_{\rm GGX}}{24}~\Big(k_{\!Y}+k_{\!W}-\frac{8}{3}~
 k_{\!C}\Big)\,.
\end{eqnarray} 
Due to Eq.~(\ref{kac}) the RHS equals zero, and with 
Eqs.~(\ref{ccx}),
(\ref{wwx}) and (\ref{yyx}) we obtain the following condition on 
the
$X$-charges
\begin{equation}
X_{H^\mcal{U}}+X_{H^\mcal{D}}+\sum\limits_{i=1}^3
\Big(X_{L^i}+X_{\ol{E^i}}-X_{Q^i}-X_{\ol{D^i}}\Big)=0\,.
\label{linkshs}
\end{equation} 
Now the low energy mass matrix of the $d$-quarks,
$\bsym{M^{(D)}}$, is $\langle H^\mcal{D}\rangle~\bsym{Y^{
(D)}}$, and the low energy Yukawa coupling matrix $\bsym{Y^{(D)
}}$ is the renormalization group evolved $\bsym{G^{(D)}}$.
Disobeying virtually all of the steps which are outlined in the next
section (to get from the string scale to the weak scale) we
approximate here
\begin{eqnarray}
\det\bsym{Y^{(D)}}=\det \bsym{\wtil{g}^{(D)}}~\cdot~
\eps^{3X_{\!H^\mcal{D}}+\sum_i (X_{\!Q^i}+X_{\ol{\!D^i}})}\,,
\end{eqnarray}
with $\wtil{g}^{(D)}_{~~~ij}\equiv{g^{(D)}}_{\!ij}~\wtil{\Theta}
[X_{Q^i}+X_{H^\mcal{D}}+X_{\ol{D^j}}]$.\footnote{Thus in order to 
have
$m_d\neq0$ one needs $\det\bsym{\wtil{g}^{(D)}}\neq0$. Using {\it
e.g.}  the simplest Ansatz for the $\wtil{g}^{(D)}_{~~ ~ij}$, namely
all of them being equal to unity, is not an option. In 
Ref.~\cite{mnr}
$\det\,\bsym{\wtil{g}^{(U)}}=0$ was used to get a massless $u$-quark,
see also footnote~$^{\ref{fu9}}$.} The same holds for the mass 
matrices of
the charged leptons.  So from Eq.~(\ref{linkshs}) we obtain
\begin{eqnarray}\label{nochnegleichung}
\eps^{X_{H^\mcal{U}}+X_{H^\mcal{D}}}~=~\Bigg[
\frac{\det\bsym{{M^{(D)}}}  }{ \det
\bsym{M^{(E)}}}\cdot\frac{\det\bsym{\wtil{g}^{(E)}}}
{ \det\bsym{\wtil{g}^{(D)}}}\Bigg]\,.
\end{eqnarray}
Assuming $\det\bsym{\wtil{g}^{(D)}}\approx\det\bsym{\wtil{g}^{(E)}}$
and using
\begin{equation} 
m_d \cdot m_s \cdot m_b~=~\det
\bsym{M^{(D)}},~~~ m_e \cdot m_\mu \cdot m_\tau~=~\det
\bsym{M^{(E)}}\,,
\end{equation}
Eq.~(\ref{nochnegleichung}) can be rewritten as
\begin{equation}
\eps^{X_{H^\mcal{U}}+X_{H^\mcal{D}}}~=~
\frac{m_d\cdot m_s \cdot m_b}{m_e\cdot m_\mu \cdot m_\tau}\;.
\end{equation}
With Eqs.~(\ref{1}), (1.2), and (1.3) we get 
\begin{eqnarray}\label{z}
X_{H^\mcal{U}}+X_{H^\mcal{D}}~~=~~~~0~~~\mbox{or}~-1\;,
\end{eqnarray}
a result that is reproduced later on in a somewhat different way, see
Table 1. Looking at Eq.~(\ref{mue}) we find that for $X_{H^\mcal{U}}+
X_{H^\mcal{D}}=0$ we have to have $|M^{(\mu)}|\ll M_{\!s}$, thus the
hierarchy problem remains.  On the other hand if
$X_{H^\mcal{U}}+X_{H^\mcal{D}}=-1$ one has that $\mu=\eps\cdot
M_{3/2}$. Hence $M^{(\mu)}$ (which is also the mass scale for the
$K_i$) is then naturally chosen to be $M_{\!s}$. As explained in more
detail below, the $K_i$ can be rotated away which however changes the
$\mu$ ({\it c.f.}\, Step 3 of the next section). So if $M^{(\mu)}\sim
M_{\!s}$, one thus needs $X_{L^i}+ X_{H^\mcal{U}}$ to be either $\geq
24$ (so that $M^{(\mu)}\sim M_{\!s}$ is adequately suppressed) or 
$<0$
and/or fractional so that the entry is forbidden.  In that way the $K_i$ do not generate an effective
$\mu$ which is too large, see Eq.~(\ref{kbeforefn}).

A similar kind of calculation was presented in Refs.~\cite{br,n}: 
The
conditions for Green-Schwarz anomaly cancellation, 
Eq.~(\ref{cancel}),
give
\begin{equation}
\frac{1}{5}~\mcal{A}_{\rm YYX}~=~\frac{1}{2}~\mcal{A}_{\rm YYX}~
+~\frac{1}{2}~\mcal{A}_{\rm WWX}~-~
\mcal{A}_{\rm CCX}\,.
\end{equation}
Expressing the RHS in terms of $X$-charges gives
\begin{eqnarray}
\frac{1}{5}~\mcal{A}_{\rm YYX}&=&X_{H^\mcal{D}}+X_{H^\mcal{U}}+
\sum_i\big(X_{L^i}+X_{\ol{E^i}}-\frac{1}{3}X_{Q^i}+\frac{1}{3}
X_{\ol{U^i}}-\frac{2}{3}X_{\ol{D^i}}  \big)\nonum\\
&=&3X_{H^\mcal{D}}+\sum_i \big(X_{L^i}+X_{\ol{E^i}}\big)~+~
\frac{1}{3}~\Big(~3~X_{H^\mcal{U}}+\sum_i \big(X_{Q^i}+
X_{\ol{U^i}}\big) ~ \Big)\nonum\\
                                  &~&~~~-~\frac{2}{3}~
\Big(~3~X_{H^\mcal{D}}+\sum_i \big(X_{Q^i}+X_{\ol{D^i}}~\big)~ 
\Big). 
\end{eqnarray}
Similarly  to the previous calculation, we get
\begin{eqnarray}
\eps^{\frac{\mcal{A}_{\rm YYX}}{5}}&=&\frac{\frac{
\displaystyle{m_e\cdot m_\mu
\cdot m_\tau}}{\displaystyle{\langle{{{H}^\mcal{D}}}\rangle^3}} 
\left(\frac{\dpst{m_u\cdot 
m_c \cdot m_t}}{\dpst{\langle {{{H}^\mcal{U}}}\rangle^3}} 
\right)^{\frac{1}{3}}}{ \left(\frac{\dpst{m_d\cdot m_s\cdot m_b}}
{\dpst{\langle
 {{{H}^\mcal{D}}}\rangle^3}} \right)^{\frac{2}{3}}}\,.
\end{eqnarray}
Using Eqs.~(1.1-1.5) and $\cot\beta=\big\langle{{{H}^\mcal{D}}}\big
\rangle/\big\langle{{{H}^\mcal{U}}} \big\rangle$ this reads
\begin{eqnarray}\label{arda}
\eps^{\frac{\mcal{A}_{\rm YYX}}{5}}&=&\frac{{m_t}^2~
\eps^{6+(0,1,2~or~3)+(0~or~1)}}{\big\langle{{{H}^\mcal{U}}}
\big\rangle^2}\,.
\end{eqnarray}
With $~~\big\langle{{{H}^\mcal{U}}}\big\rangle^2+\big\langle{{{H}
^\mcal{D}}} \big\rangle^2~=~(246~\mbox{GeV})^2~~$ one arrives at 
\begin{eqnarray}
\eps^{\frac{\mcal{A}_{\rm YYX}}{5}}&=&~\eps^{6+(0,1,2~or~3)+(0~or~1)}
~\frac{1+\tan^2\beta}{\tan^2\beta}\left(\frac{m_t}{246~\mbox{GeV}}
\right)^2\;.
\end{eqnarray}
Experimentally, the parameters are restricted by \cite{lephiggs,pdg} 
\begin{equation}
2.4\leq \tan\beta\leq 50~~~\mbox{and}~~~m_t=(174.3\pm5.1)~\mbox{GeV},
\end{equation}
and therefore ${{\mcal{A}_{\rm YYX}}/{5}}\sim{6+(0,1,2~or~3)+(0~or~1)
}$.  It follows that $\mcal{A}_{\rm YYX}\!=\!0$ is impossible, and
hence ({\it c.f.} Eq.~(\ref{cancel})) $\mcal{A}_{\rm WWX}\!=\!
\mcal{A}
_{\rm CCX}\!=\!0$ are also not possible. So $U(1)_X$ has to be
anomalous, and we need the Green-Schwarz mechanism, which fixes the
breaking of $U(1)_X$ to be close to the string scale.

Further phenomenological constraints on the $X$-charges arise more
systematically from requiring the quark mass matrices to reproduce
Eqs.~(1.3), (\ref{4}), (\ref{5}), (\ref{NeWW}) and (\ref{6}) at high 
energies. There
are four possible pairs of $\bsym{G^{(U)}}$ and $\bsym{G^{(D)}}$ (in
Appendix~\ref{illustration} the validity of these mass matrices is
illustrated):
\begin{eqnarray}\label{oins}
&\bsym{G^{(U)}}~\propto~
\left(\begin{array}{lll}
\eps^8 & \eps^{5\phantom 3} & \eps^{3} \\
\eps^{7\phantom 3} & \eps^4 & \eps^2 \\
\eps^5 & \eps^2 & 1 \end{array}\right),~~~&
{\bsym{G^{(D)}}}~\propto~
\left(\begin{array}{lll}
\eps^{4\phantom 3} & \eps^{3\phantom 3} & \eps^3 \\
\eps^3 & \eps^2 & \eps^2 \\
\eps & 1 & 1 \end{array}\right),\\
\label{zwo}
&{\bsym{G^{(U)}}}~\propto~\left(\begin{array}{lll}
\eps^8 & \eps^{5\phantom 3} & \eps^3 \\
\eps^{13} & \eps^4 & \eps^2 \\
\eps^{11} & \eps^2 & 1 \end{array}\right),~~~&
{\bsym{G^{(D)}}}
~\propto~
\left(\begin{array}{lll}
\eps^{4\phantom 3} & \eps^{3\phantom 3} & \eps^3 \\
\eps^9 & \eps^2 & \eps^2 \\
\eps^7 & 1 & 1 \end{array}\right),\\
\label{droi}
&{\bsym{G^{(U)}}}~\propto~\left(\begin{array}{lll}
\eps^8 & \eps^{6\phantom 3} & \eps^4 \\
\eps^{6\phantom 3} & \eps^4 & \eps^2 \\
\eps^4 & \eps^2 & 1 \end{array}\right),~~~&
{\bsym{G^{(D)}}}
~\propto~
\left(\begin{array}{lll}
\eps^{4\phantom 3} & \eps^{4\phantom 3} & \eps^4 \\
\eps^2 & \eps^2 & \eps^2 \\
1 & 1 & 1 \end{array}\right),\\
\label{fihr}
&{\bsym{G^{(U)}}}~\propto~\left(\begin{array}{lll}
\eps^8 & \eps^{6\phantom 3} & \eps^4 \\
\eps^{14} & \eps^4 & \eps^2 \\
\eps^{12} & \eps^2 & 1 \end{array}\right),~~~&{\bsym{G^{(D)}}}
~\propto~
\left(\begin{array}{lll}
\eps^{4\phantom 3} & \eps^{4\phantom 3} & \eps^4 \\
\eps^{10} & \eps^2 & \eps^2 \\
\eps^8 & 1 & 1 \end{array}\right).
\end{eqnarray}
The matrices in Eqs.~(\ref{oins}, \ref{zwo}) were suggested in
Refs.~\cite{dps,blr} and more rigorously derived in
Ref.~\cite{cl}.\footnote{\label{nrneun}Utilizing Eq.~(\ref{bilara}),
the matrices in Eq.~(\ref{zwo}) can be generated from naive exponents
$X_{{Q^i}}+X_{H^\mcal{U}}+X_{\ol{U^j}}$ and $X_{{Q^i}}+X_{H^\mcal{D}}
+X_{\ol{D^j}}$ given as follows, respectively:
\begin{eqnarray}
\left(\begin{array}{rrr}
8 & -1 & -3 \\ 13 & 4 & 2 \\ 11 & 2 & 0
\end{array}\right)_{\!ij},~~~\left(\begin{array}{rrr} 4 & {-3} & {-3}
\\ 9 & 2 & 2 \\ 7 & 0 & 0
\end{array}\right)_{\!ij}.\nonum
\end{eqnarray} 
The negative exponents give textures which are filled up in the
process of canonicalization of the K\"ahlerpotential, see Step~2 of
the next section.} They are in accord with the {\it lefthand} choice
of Eq.~(1.7). Ref.~\cite{haba} pointed out that the {\it righthand}
choice of Eq.~(\ref{6}) can be achieved from the mass matrices in
Eq.~(\ref{droi}).\footnote{Note that ${\bsym{G^{(U)}}}$ in
Eq.~(\ref{droi}) is the only matrix that is compatible with an
$SU(5)$-GUT: ${\bsym{G^{(U)} }}$ is symmetric, being in accord with
the $SU(5) $-requirement $X_{\ol{U^i}}=X_{Q^i}\,$. To have 
furthermore
$X_{L^i}=X_{\overline{D^i}},~X_{\overline{E^i}}=X_{Q^i}$ one needs
$X_{L^1}=X_{L^2}=X_{L^3},~y=1,~z=0$, in the notation below.}  For
completeness' sake, in analogy to Eq.~(\ref{zwo}) we found here that
the matrices in Eq.~(\ref{fihr}) are also in accord with the 
righthand
choice of Eq.~(1.7).\footnote{\label{nrelf}Utilizing
Eq.~(\ref{bilara}), the matrices in Eq.~(\ref{fihr}) can be generated
from naive exponents $X_{{Q^i}}+ X_{H^\mcal{U}}+X_{\ol{U^j}}$ and
$X_{{Q^i}}+ X_{H^\mcal{D}}+X_{\ol{D^j}}\,$ given as follows,
respectively:
\begin{eqnarray}
\left(\begin{array}{rrr}
8 & {-2} & {-4} \\
{14} & 4 & 2 \\
{12} & 2 & 0 \end{array}\right)_{\!ij},~~~\left(\begin{array}{rrr}
4 & {-4} & {-4} \\
{10} & 2 & 2 \\
8 & 0 & 0 \end{array}\right)_{\!ij}\nonum.
\end{eqnarray}
As in 
footnote~$^{\ref{nrneun}}$, the negative exponents give 
textures which are filled up in 
the process of canonicalization of the K\"ahlerpotential, see 
Step~2 of 
the next section.} 

It is convenient to parameterize the $X$-charges in terms of the
entries of the matrices in Eqs.~(\ref{oins}-\ref{fihr}):
\begin{eqnarray}\label{rrr}
X_{Q^i}+X_{H^\mcal{U}}+X_{\ol{U^j}}&=&\left(\begin{array}{rrr}
8 & 5+y  &  3+y        \\
7-y &4  & 2           \\
5-y & 2  & 0     \end{array}\right)_{\!\!ij}~~,\\
\nonum\\
\label{RRR}X_{Q^i}+X_{H^\mcal{D}}+X_{\ol{D^j}}&=&\left(
\begin{array}{rrr}
4+x & 3+y+x  &  3+y+x        \\
3-y+x &2+x  & 2+x             \\
1-y+x & x & x      \end{array}\right)_{\!\!ij}~~,
\end{eqnarray} 
with 
\begin{equation}
x=0,1,2,3\quad {\rm if}\quad y=-7,0,1\quad {\rm and} 
\quad x=0,1,2\quad 
{\rm if} \quad y=-6\,. 
\label{models}
\end{equation}
A negative $y$ causes some naive exponents in $\boldsymbol{G^{(U)}},
\boldsymbol{G^{(D)}}$
to be negative as well (see footnotes \ref{nrneun}
and \ref{nrelf}) and is thus forbidden; these textures are then 
filled
up in the process of canonicalizing the K\"ahlerpotential. In 
addition
to Eqs.~(\ref{rrr}, \ref{RRR}), Eq.~(\ref{1}) motivates the
parameterization
\begin{equation}\label{lll}
X_{L^i}+X_{H^\mcal{D}}+X_{\ol{E^i}}~=~\left(\begin{array}{r}
4+z+x\\
2+x \\
x \end{array}\right)_{\!\!i}\;,
\end{equation}
with $z=0,1$. Finally, Eq.~(1.2) leads to
\begin{equation}\label{LLL}
X_{L^3}+X_{\ol{E^3}}=X_{Q^3}+X_{\ol{D^3}}\;.
\end{equation}
Solving Eqs.~(\ref{rrr}-\ref{LLL}) and the conditions for
Green-Schwarz anomaly cancellation,  {\it i.e.} Eqs.~(\ref{cancel},
\ref{ccx}-\ref{yxx}) for the $X$-charges leads to the
expressions given in Table~1; the free parameters chosen here are 
$x,y
,z,X_{{L^1}}, X_{{L^2}},X_{{L^3} }$.\footnote{\label{soertien}
 The reasons 
we have
chosen to work with $X_{{L^1}}, X_{{L^2}},X_{{L^3}}$ are as follows:
\emph{(i)}  The quark sector is fairly well known, so 
Froggatt-Nielsen
model-building does not need an $X$-charge of a quark to be an
adjustable parameters.
\emph{(ii)} If $\boldsymbol{G^{(E)}},\boldsymbol{G^{(N)}}$ and 
$\boldsymbol{\Gamma}$ are without supersymmetric zeroes, one 
finds that the 
Maki-Nagakawa-Sakata matrix \cite{mns} is approximately given by 
$${U^{M\!N\!S}}_{ij}\sim\epsilon^{|X_{L^i}-X_{L^j}|},~~~
\mbox{just like}~~~ 
{U^{C\!K\!M}}_{ij}~~\sim~~ \epsilon^{|X_{Q^i}-X_{Q^j}|};$$ 
furthermore (see Ref. \cite{blr,eir})
$$\boldsymbol{G^{(N)}}\cdot\boldsymbol{\Gamma}^{-1}\cdot
\boldsymbol{G^{(N)}}^T \sim~\frac{\epsilon^{X_{L^i}+X_{L^j}+
2X_{H^\mathcal{U}}}}
{M^{(\overline{N})}}\;.$$}$^,$\footnote{If one believes that 
the actual expansion parameter is $\epsilon^2$ rather than 
$\epsilon$, one has to work with $x=0,2$, $~y=1,-7$, $~z=0$.}  
\begin{figure}
\begin{center}
\vspace{-1.3cm}
\begin{tabular}{|rcl|}
\hline
$\phantom{\Bigg|}X_{H^\mcal{D}}$&$=$&$-~\frac{1}{54+9x+6z}
\big[18 + X_{L^1}
(18+2x+5z) + 
        X_{L^2}(12 + 2x + 2z)$\\
 & & $~~~~~+ X_{L^3}(6 + 2x + 2z)- x(36+ 6x+5z) - 18y- 2z^2\big]$\\
$\phantom{\Bigg|}X_{H^\mcal{U}}$&$=$&$-X_{H^\mcal{D}}-z $\\
$\phantom{\Bigg|}X_{Q^1}$&$=$&$\frac{1}{9}\big[30 - (X_{L^1} 
+ X_{L^2} + 
X_{L^3}) + 3x + 6y + 4z\big]$\\
$\phantom{\Bigg|}X_{Q^2}$&$=$&$X_{Q^1}-1-y $\\
$\phantom{\Bigg|}X_{Q^3}$&$=$&$X_{Q^1}-3-y $\\
$\phantom{\Bigg|}X_{\ol{U^1}}$&$=$&$X_{H^\mathcal{D}}-X_{Q^1}+8+z $\\
$\phantom{\Bigg|}X_{\ol{U^2}}$&$=$&$X_{\ol{U^1}}-3+y $\\
$\phantom{\Bigg|}X_{\ol{U^3}}$&$=$&$X_{\ol{U^1}}-5+y $\\
$\phantom{\Bigg|}X_{\ol{D^1}}$&$=$&$-X_{H^\mathcal{D}}-X_{Q^1}+4+x 
$\\
$\phantom{\Bigg|}X_{\ol{D^2}}$&$=$&$X_{\ol{D^1}}-1+y $\\
$\phantom{\Bigg|}X_{\ol{D^3}}$&$=$&$X_{\ol{D^1}}-1+y $\\

$\phantom{\Bigg|}X_{\ol{E^1}}$&$=$&$-X_{H^\mathcal{D}}+4 - X_{L^1} 
+ x + z $\\
$\phantom{\Bigg|}X_{\ol{E^2}}$&$=$&$-X_{H^\mathcal{D}}+2- X_{L^2}+x  
$ \\ 
$\phantom{\Bigg|}X_{\ol{E^3}}$&$=$&$-X_{H^\mathcal{D}}\phantom{+2}~- 
X_{L^3}+x  $ 
\\\hline
\end{tabular}
$~$\\
Table 1: $~\phantom{\Bigg|}$  The constrained $X$-charges.
\label{table1111}
\end{center}
\end{figure}
In the Table in Appendix~A we have translated all the models given
in Refs.~[1-16] into the (standard) notation above. This should allow
for an easier comparison. Note that one has 
\begin{equation}\label{ccx(xz)}
\mcal{A}_{\rm CCX}=3\, (6 + x + z)\;,
\end{equation}
as was actually already apparent in Eq.~(\ref{arda}), and
\begin{eqnarray}
X_{L^i}+X_{H^\mcal{D}}+X_{\ol{E^j}}~=~\left(
\begin{array}{rrr}
4+x+z & 2+x  & x        \\
4+x+z & 2+x  & x              \\
4+x+z & 2+x  & x      \end{array}\right)_{\!\!ij}~+~X_{L^i}-X_{L^j}.
\end{eqnarray}

As an example, consider the first model of Ref.~\cite{blr}. The
charges are determined by $x=2,~y=0,~z=0,~X_{\ol{L^1}}=-12,~
X_{\ol{L^2}}=
-13,~X_{\ol{L^3}}=55$. Thus one has, using Table~1
\begin{equation}
\begin{tabular}{|c|} \hline~~ $X_{H^\mcal{D}}=0,\phantom{\Big|}~~
X_{H^\mcal{U}}=0~~$\\\hline\end{tabular}\nonum
\end{equation}
and the quark and lepton charges

\vspace{0.5cm}

\begin{center}
\begin{tabular}{|c|c|c|c|c|c|}
\hline
\bf{Generation}\phantom{$\Big|$} $\bsym{i}$~ &~~~ $
\bsym{X_{\!Q^i}}$~~~ &~~~ $\bsym{X_{\!\ol{D^i}}}$~~~ &~~~ $
\bsym{X_{\!\ol{U^i}}}$~~~ &~~~ $\bsym{X_{\!L^i}}$~~~ &~~~ $
\bsym{X_{\!\ol{E^i}}}$~~~       \\\hline
1\phantom{\Bigg|} & $~~\frac{2}{3}$ &$\frac{16}{3}$ &$\frac{22}{3}$ 
&$-12\phantom{-}$ 
&$~~18$  \\ \hline
2\phantom{\Bigg|} & $-\frac{1}{3}\phantom{-}$ &$\frac{13}{3}$ 
&$\frac{13}{3}$ &$-13\phantom{-}$ 
&$~~17$  \\\hline 
3\phantom{\Bigg|} & $-\frac{7}{3}\phantom{-}$ &$\frac{13}{3}$ 
&$\frac{7}{3}$ &$55$ 
&$-53\phantom{-}$  \\ \hline 
\end{tabular}~.
\end{center}$~$
 
This results in the quark mass matrices which are given in
Eq.~(\ref{oins}), and (taking into account supersymmetric zeros)
the leptonic mass matrix
\begin{eqnarray} 
\bsym{G^{(E)}}\sim\left(\begin{array}{lll}\eps^6  &\eps^5 & 
0 \\\eps^5 &\eps^4 &0 \\\eps^{73} &\eps^{72} &\eps^2    
\end{array}\right)\,.
\end{eqnarray}

%
%
\section{\label{procedure}The Procedure}
\setcounter{equation}{0}
Taking a top-down approach we start off at or close to the string
scale with a family symmetry, {\it i.e.} with a certain $X$-charge
assignment for the left-chiral superfields. We determine the 
exponents
of $\eps$ of all allowed interactions, including the baryon- and
lepton-number violating interactions. We thus also determine the
supersymmetric zeroes.  We then translate this into a low-energy 
model
in order to compare with the measured fermion masses, their mixings
and especially the experimental bounds on the barity- and 
lepton-parity
violating coupling constants.  To do so, we perform the following
steps:
\begin{enumerate}

\item In models that contain right-handed neutrinos, the 
corresponding
 scalar superpartners can acquire VEVs which are determined by the
 minima of the scalar potential. The scalar potential is non-negative
 (apart from supergravity effects) and supersymmetry has not been
 broken yet (since we are just below the $U(1)_X$ breaking scale, 
thus
 quite close to the string scale), so the absolute minima of the
 scalar potential are given by the roots of the scalar
 potential. These roots in turn are given by the roots of the
 superpotential.\footnote{\label{six10}Of course this consideration
 neglects the part of the scalar potential which arises from the
 $D$-terms.  However, this is justified by our starting point: We are
 working with \emph{one} flavon field, {\it i.e.} $U(1)_X$ gets 
broken
 predominantly by the scalar component of $A$ acquiring a VEV, rather
 than the scalar components of the $\overline{N^i}$ acquiring
 VEVs. This is equivalent to $\upsilon\gg \Delta^i$, which is
 justified with hindsight by the $\Delta^i$ being functions of
 $\Xi_i,\Gamma_{ij}$.}  Finding the roots of $\mcal{W}$ is equivalent
 to finding the linear constant shifts
\begin{eqnarray}
\ol{N^i}(x,\theta)\longrightarrow\ol{N^i}(x,\theta)+\Delta^i,
\end{eqnarray}
such that the tree-level tadpole term $\Xi_i~\ol{N^i}$ in the
superpotential vanishes\footnote{$\Xi_i$ is not regenerated by the
RG-flow after having been shifted away, see the expression in
Ref.~\cite{mv} for the beta-function of a general linear term of a
superpotential.} [note that the Giudice-Masiero mechanism might
produce, $\Xi_i\sim M_{3/2}M_{\!s}$, {\it c.f.} Eq.~(\ref{xigm})]. 
This
shift results in constant and thus harmless terms (ignoring the
problem of the cosmological constant) in the K\"ahler- and
superpotential, as well as the redefinitions of the superpotential
coupling constants:
\begin{eqnarray}\label{elf2}
K_i&\longrightarrow&K_i~+~{G^{(N)}}_{\!ij}~\Delta^j,\nonum\\
\mu&\longrightarrow&\mu~+~\Upsilon_i~\Delta^i,\nonum\\
\Xi_i&\longrightarrow&\Xi_i~+~2~\Gamma_{ij}~\Delta^j~+~3~{
\Lam^{\prime
\prime\prime}}_{ijk}~\Delta^j~\Delta^k~\stackrel{!}{=}~0,\nonum\\
\Gamma_{ij}&\longrightarrow&\Gamma_{ij}~+~3~{\Lam^{\prime
\prime\prime}}_{ijk}~\Delta^k.
\end{eqnarray} 
It is important to notice that the equations above lead to a mixing
among the coupling constants.

\item The breaking of $U(1)_X$ generates a  K\"ahlerpotential
which does not have the canonical form \cite{lns1}. Thus we must
perform a transformation of the relevant superfields to the canonical
basis \cite{fkw}. For example, for the quark doublets we obtain for
the relevant K\"ahlerpotential term (where the ${H^{(Q)}}_{\!ij}$ are
K\"ahlerpotential coupling constants which are to be canonicalized)
\begin{eqnarray}
\label{st}
\ol{Q^i}~{H^{(Q)}}_{\!ij}~Q^j&=&\ol{\Big[\bsym{
\sqrt{\bsym{D_{\!H^{(Q)}}}}~
\bsym{U_{\!H^{(Q)}}}~Q}\Big]^i}~\delta_{ij}~\Big[\bsym
{\sqrt{\bsym{D_{\!H^{(Q)}}}}~\bsym{U_{\!H^{(Q)}}}~Q}\Big]^j.
\end{eqnarray}
$\bsym{D_{\!H^{(Q)}}}$ is a diagonal matrix, its entries are the
eigenvalues of the Hermitian matrix $\bsym{H^{(Q)}}$; the unitary
matrix $\bsym{U_{\!H^{(Q)}}}$ performs the diagonalization. We define
the matrix
\begin{equation}
\bsym{C^{(Q)}}\!
\equiv \!\sqrt{\bsym{D_{\!H^{(Q)}}}}~ \bsym{U_{\!H^{(Q)}}}\,.
\label{**}\end{equation}
The transformation to the canonical basis can then be written as
\begin{equation}
Q^i\rightarrow {C^{(Q)}}^i_{~j}~Q^j.
\end{equation}
The redefinition also affects the superpotential, {\it e.g.} the
$u$-type quark Yukawa couplings
\begin{equation}\label{gupost}
 \bsym{G^{(U)}}\longrightarrow\frac{1}{\sqrt{H^{\left(H^\mcal{U}
\right)}}}~{{\bsym{C^{(Q)}}}^{-1}}^T~\bsym{G^{(U)}}~ {
\bsym{C^{(\ol{U})}}}^{-1}.
\end{equation}
In this way the canonicalization of the K\"ahlerpotential 
``fills up''
possible supersymmetric zeroes (which are due to holomorphy and
 fractional $X$-charges \cite{lns1}),\footnote{There are exceptions though: If one
works with certain $X$-charge assignments that involve fractional
charges ({\it e.g.}  the one stemming from $X_{\ol{D^3}}=\frac{1}{4}$
and $X_{H^\mcal{D}}=\frac{1}{2}$) the four zeroes of $\bsym{G^{(E)}}$
are not filled up. This seems to have first been mentioned in
Ref.~\cite{cck}.} in this particular case in the mass matrix $
\bsym{G^{(U)}}$.

Since $L^i$, $H^\mcal{D}$ have the same gauge quantum numbers under
$SU(3)_C\times SU(2)_W\times U(1)_Y$, they can be mixed by kinetic
terms. So the canonicalization of the K\"ahler\-potential causes
mixing of ${G^{(E)}}_{\!ij}$ with $\Lam_{ikj}$, ${G^{(D)}}_{\!ij}$
with ${\Lam^\prime}_{ikj}$ (an example is given at the end of this
section), $\mu$ with $K_i$, ${G^{(N)}}_{\!ij}$ with $\Upsilon_{i}$,
and $\Psi_{ijkl}$ with ${\Psi^\prime}_{ijk}$.

To maintain the canonicalized form, all further field redefinitions
have to be unitary.\footnote{Actually the canonicalization is 
affected
by loop-diagrams, but these effects can be neglected because they are
small: {\it e.g.} the graphs contributing to the propagators of the
neutrinos are similar to the graphs that radiatively generate tiny
neutrino masses.}  To demonstrate an exact canonicalization, consider
the quark doublets with the $U(1)_X$ flavour charges $X_{Q^1}\!=\!4,~
X_{Q^2}\!=\!2,~X_{Q^3}\!=\!0$ and setting ${h^{(Q)}}_{ij}=1$ ({\it
c.f.} Eq.~(\ref{hq})). For the K\"ahlerpotential matrix of the $Q^i
$-superfields we obtain
\begin{equation}\label{Hq}
\bsym{H^{(Q)}}=\left(\begin{array}{ccc}
1 & \eps^2 & \eps^4 \\
\eps^2 & 1 & \eps^2 \\
\eps^4 & \eps^2 & 1 \end{array}\right).
\end{equation}
This is canonicalized by the matrix ({\it c.f.} Eq.~(\ref{**})), 
\begin{eqnarray}\label{exactamente}
{{\bsym{C^{(Q)}}}^{-1}}^\dagger=\frac{1}{\sqrt{1-\eps^4~}}
\left(\begin{array}{ccc}-\frac{1}{\sqrt{2}} & 0 &\frac{1}{\sqrt{2}}\\
& & \\
\frac{w_+}{2} & \frac{w_+f_-}{2} & \frac{w_+}{2} \\ & & \\
\frac{w_-}{2} & \frac{w_-f_+}{2} & \frac{w_-}{2} \end{array}\right),
\end{eqnarray}
with
$w_\pm=\sqrt{1\pm\frac{3\eps^2}{\sqrt{8+\eps^4~}}~},~f_\pm
=\frac{3\eps^2\pm\sqrt{8+\eps^4~}}{2+\eps^4\pm\eps^2
\sqrt{8+\eps^4~}}$. As $\eps$ is a small quantity, 
the expression above can be approximated to leading order in $\eps$ 
as
\begin{eqnarray}\label{cq}
{\bsym{C^{(Q)}}^{-1}}^\dagger\approx\left(\begin{array}{ccc}
-\frac{1}{\sqrt{2}} & 0 &\frac{1}{\sqrt{2}}\\
& & \\
\frac{1}{2}+\frac{3~\eps^2}{8\sqrt{2}} & -\frac{1}{\sqrt{2}}
-\frac{5~\eps^2}{8} & \frac{1}{2}+\frac{3~\eps^2}{8\sqrt{2}} 
\\ & & \\ \frac{1}{2}-\frac{3~\eps^2}{8\sqrt{2}} & \frac{1}{\sqrt{2}}
-\frac{5~\eps^2 }{8} &  \frac{1}{2}-\frac{3~\eps^2 }{8\sqrt{2}} 
\end{array}\right).
\end{eqnarray}
However, while this article was reviewed, a preprint \cite{jj} 
came out, 
stating and nicely demonstrating that in many cases 
one can utilize a further
 unitary 
transformation to bring ${\bsym{C^{(Q)}}^{-1}}^\dagger$ to its 
``standard'' form, see Eq.~(\ref{bilara}).

\item Next we perform a unitary field redefinition of the fields 
$\left(\!\!\begin{array}{cc}H^\mcal{D},L^i\end{array}\!\!\right)$ 
such
that the term $K_iL^iH^\mcal{U}$ vanishes.  This was first realized 
in
Ref.~\cite{hs}, however, the rotation matrix was approximate and in
addition only valid for real $\mu$ and $K_i$, with $K_i/\mu\ll1$.  We
have implemented the full transformation for {\it arbitrary complex}
$K_i,\mu$.  With $K\equiv\sqrt{{K_i}^*K^i~}$ and $\mcal{M}\equiv\sqrt
{\mu^*\mu+K^2~}$ (summation over repeated indices implied) the exact
transformation is given by
\begin{eqnarray}\label{lhu}
\nonum\\
\underbrace{\frac{|\mu|}{\mcal{M}}\left(\begin{array}{ccc}
1 &~ & \Big(\!\frac{K_j}{\mu}\!\Big)^{\!*}\\
 & ~ & \\
-\frac{K^i}{\mu} &~& \frac{{K_j}^*~K^i}{K^2}\Big(\!1\!-\!\frac
{\mcal{M}}{|\mu|}\Big)\!+\!\frac{\mcal{M}}{|\mu|}~\delta^
{^i}_{^{~\!_j}}\end{array}\right)}_{\equiv~{ {\displaystyle
\bsym{{U}_{\!\!\not K_i}^{\large{~\ast}}}}}}\cdot\left(
\begin{array}{c}
\mu \\~\\
K^j\end{array}\right)
&=&\left(\begin{array}{c}
\frac{\mu}{|\mu|}\mcal{M} \\~\\
0\end{array}\right).~~~~~\\\nonum
\end{eqnarray}
The $SU(4)$ matrix acting on the fields $\left(\!\!\begin{array}{cc}
H^\mcal {D}, L^i\end{array}\!\!\right)$ is given by $\bsym{{U}_{\!\!
\not K_i}}^{T}$ (we shall refer to it as a rotation matrix, since for
real coupling constants it is an $S\!O\!(4)$ matrix). The rotation
gives further mixing of the coupling constants: ${G^{(E)}}_{\!ij}
\lra{ \Lam}_{ikj}$, ${G^{(D)}}_{ij}\lra{\Lam ^\prime}_{ikj}$,
${G^{(N)}}_{\!ij}\lra\Upsilon_{i}$ and $\Psi_{ijkl}\lra{\Psi^\prime}
_{ijk}$.  As explained in Section \ref{FroNie}, if $m_e/m_\mu\!
\approx\!\eps^5$, one naturally has $M^{(\mu)}\sim M_{\!s}$. So one 
obtains the phenomenologically unacceptable relation that $\frac{\mu}
{|\mu|}\mcal{M}$ is much larger than the weak scale, unless the $K_i$
are suppressed by at least $\eps^{X_{L^i}+X_{H^\mcal{U}}}\!\sim\!\eps
^{24}$ ($\eps^{24}M_{\!s}\sim$ weak scale), see the text below
Eq.~(\ref{z}).

\item In the next step, the coupling constants are evolved 
according  
to their renormalization group equations (RGEs) down from the $U(1)_X
$-breaking scale to the weak scale. The one-loop RGEs of the $\not\!
\!R_p$ coupling constants are given in Ref.~\cite{bbpw}, the
two-loop RGEs of the gauge coupling constants and the coupling
constants of the superpotential of the $\not\!\!R_p$-$M\!S\!S\!M$
 were
derived in Ref.~\cite{add}, the one-loop RGEs of the soft
supersymmetry-breaking coupling constants of the $\not\!\!R_p$-$M\!S
\!S\!M$ are given in Ref.~\cite{add3}, see also Ref.~\cite{cw} for 
the third generation only. The RG-flow once again causes ${G^{(E)}}_
{\!ij}\lra\Lam_{ikj}$, ${G^{(D)}}_{\!ij}\lra{\Lam^\prime}_{ikj}$
and ${G^{(N)}}_{\!ij}\lra\Upsilon_{i}$ mixing and regenerates $K_i$
\cite{nardi}. Likewise for the corresponding soft coupling constants.
  
\item The superfields $L^i$ and $H^\mcal{D}$ are rotated unitarily 
such 
that only the scalar neutral $CP$-even components of $H^\mcal{D}$ and
$H^\mcal{U}$ acquire VEVs. The relevant matrix is given by
Eq.~(\ref{lhu}), with the appropriate replacements.  The rotation
again mixes the regenerated $K^i$ with $\mu$, and again causes 
${G^{(E
)}}_{\!ij}\lra\Lam_{ikj}$, ${G^{(D)}}_{\!ij}\lra{\Lam^\prime}_{i
kj}$, ${G^{(N)}}_{\!ij}\lra\Upsilon_{i}$ and  $\Psi_{ijkl}\lra{\Psi^
\prime}_{ijk}$ mixing.
 
\item In order to compare with experiment, we must transform into 
the 
mass basis of the quarks, which once again skews the coupling
constants. As we will assume massless neutrinos here, we do not have
to be concerned about the mass basis of the leptons.  The inclusion 
of
neutrino masses will be discussed elsewhere.\footnote{The neutrino
mass matrix pattern required for the solar and atmospheric neutrino
problems can also be obtained via the idea of Froggatt and Nielsen,
see {\it e.g.} Ref.~\cite{scheich}.}

\item Finally, in the last step, the tree and loop contributions to 
the mass matrix of the neutral fermions are determined. The
neutralinos, and the regenerated operators $K_iL^iH^\mcal{U}$ can
cause non-zero neutrino masses. The $\not\!\!R_p$ operators allow 
loop
graphs that radiatively generate masses for the neutrinos, see {\it
e.g.} Refs. \cite{gh,dl1,dl2,dl3}.\footnote{The corrections to the
neutrino masses are totally negligible if the model contains
right-handed neutrinos, in the same way as the corresponding
corrections are unimportant for the masses of the charged fermions.} 
In this paper we have not taken into account phenomenological
constraints on the $X$-charges that arise from the neutrino sector.
 
\end{enumerate}
Note that if {\it all} of the (highly undesired) couplings ${\Lam^
{\prime\prime}}_{ijk}$ are initially forbidden due to naive 
fractional
and/or negative $X$-charges, none of the above steps can regenerate
them.

Of course, performing all these steps entirely satisfactory requires
the precise knowledge of the unknown {\it a priori} arbitrary
$\mcal{O}\!(1)$ coupling constants ${g^{(U)}}_{\!ij},
{h^{(U)}}_{\!ij},
C_{H^\mcal{D}H^\mcal{U}}$, etc. Thus the only thing we can do is
assume that there is no fine-tuning leading to accidental
cancellations and to perform order-of-magnitude estimates. We would
now like to point out several caveats and also comment on how our
procedure differs from relevant previous work:
\begin{enumerate}

\item Step 1 has been neglected in the past; we have taken it into 
account whenever $X_{\ol{N^1}},X_{\ol{N^2}},X_{\ol{N^3}}$ were given
(this is not the case in the model presented in Ref.~[10], the  
models 
presented in Ref.~[12] and the model in Ref.~[16]).
The
right-handed neutrino charges did however not change whether the 
model under
consideration is compatible or not: the $X_{\overline{N ^i}}$ in 
model
[13] are fractional and thus no $\Xi_i\overline{N^i}$ is generated,
the models [14] and [15] are already ruled out even if the explicitly
given $X_{\overline{N^i}}$ are not considered, likewise for the 
models
[17]-[21].

\item For Step 2, in the past the approximation
\begin{equation}\label{bilara}
{{C^{(Q)}}^{-1}}_{\!ij}\sim\eps^{|X_{Q^i}-X_{Q^j}|}\;
\end{equation}
was utilized (likewise for the other fields), see
Ref.~\cite{blr}.\footnote{This result can be reached as follows:
Ignoring the fact that $|X_{Q^i}-X_{Q^j}|$ could be fractional,
\begin{equation}
\sum \limits_{i,j}\ol{Q^i}~ {H^{(Q)}}_{\!ij}~Q^j\approx
\sum\limits_{i,j}\ol{Q^i}~\eps^{|X_{Q^i}-X_{Q^j}|}~Q^j\,,
\nonum
\end{equation}
see Eq.~(\ref{Hq}). Now $|X_{Q^i}-X_{Q^j}|\leq
|X_{Q^j}-X_{Q^k}|+|X_{Q^k}-X_{Q^i}|$, from which follows that
$\eps^{|X_{Q^i}-X_{Q^j}|}\geq\eps^{|X_{Q^j}-X_{Q^k}|}$
$\eps^{|X_{Q^k}-X_{Q^i}|}$, and hence, neglecting prefactors of
$\mcal{O}\!(1)$, $\eps^{|X_{Q^i}-X_{Q^j}|}
\approx$ $\sum\limits_k\eps^{|X_{Q^j}-X_{Q^k}|}~
\eps^{|X_{Q^k}-X_{Q^i}|}$.
Thus 
\begin{equation}
\sum\limits_{i,j}\ol{Q^i}~\eps^{|X_{Q^i}-X_{Q^j}|}
Q^j\approx~\sum\limits_{i,j,k}\ol{Q^i}~
\eps^{|X_{Q^j}-X_{Q^k}|}\eps^{|X_{Q^k}-X_{Q^i}|}~Q^j\,,
\nonum
\end{equation}
and hence, comparing with Eq.~(\ref{st}), $\eps^{|X_{Q^i}-X_{Q^j
}|}$ $\approx{C^{(Q)}}_{\!ij}$. Furthermore, the $\eps^{|X_{Q^i}
-X_{Q^j}|}$ form a matrix that is approximately equal to its square
and thus approximately equal to its cube, so that this matrix is
approximately equal to its own inverse, so that
\begin{equation}
\Big[{C^{(Q)}}^{-1}\Big]_{\!ij}\approx\eps^{|X_{Q^i}-X_{Q^j}|}
=\eps^{|X_{Q^j}-X_{Q^i}|}\approx\Big[{{{C^{(Q)}}^{-1}}}^T\Big]_{\!ij}
=\Big[{{{C^{(Q)}}^{-1}}}^\dagger\Big]_{\!ij}\,. \nonum
\end{equation}}
For example, for the choice of quark charges $X_{Q^1}\!=\!4$, 
$X_{Q^2}
\!=\!2$ and $X_{Q^3}\!=\!0$, the transformation matrix is given by
\begin{equation}\label{istdaslebenschoen}
{\bsym{C^{(Q)}}}^{-1}=\left(\begin{array}{ccc} 1 & \eps^2 &
\eps^4 \\
\eps^2 & 1 & \eps^2 \\
\eps^4 & \eps^2 & 1 \end{array}\right).
\end{equation}
This is in gross disagreement with the exact result in 
Eq.~(\ref{cq}).
Of course, Eq.~(\ref{cq}) relied on a fine-tuning (all ${h^{(Q)}}_{\!
ij}$=1), however it illustrates that Eq.~(\ref{bilara}) might be a
very poor approximation. The correct matrix should be found somewhere
between these two extreme cases (see however the text below
Eq.~(\ref{cq}) and Ref.~\cite{jj}).  Keeping in mind that the
canonicalization of the K\"ahlerpotential is very important for
filling up supersymmetric zeroes, one sees that Eq.~(\ref{cq}) skews
the coupling constants much more than Eq.~(\ref{bilara}) and thus
leads to a much stronger unwanted attenuation of possible
hierarchies. In this paper we shall utilize Eq.~(\ref{bilara}), being
the more conservative approach.

\item Step 3 has been explicitly mentioned in the past only by 
Refs.~\cite{jvv,mnrv}. We utilize an approximation to leading order 
in
$\eps$.

\item For the time being, we shall use the most up-to-date 
GUT-scale bounds on the single $\not\!\!R_p$ coupling constants given
in Ref.~{\cite{add2}}. For products of two $\not\!\!R_p$ Yukawa
coupling constants we use the weak scale bounds\footnote{This 
approach
is conservative, as the bounds get tighter when going to higher
scales.} of Refs.~\cite{add2,dpt,dpt2,sv} (instead of using some of
the bounds of Ref.~\cite{dpt} one might wish to use the bounds of
Ref.~\cite{ayreck}: they are stricter, however less conservative). 
The
bounds are compared in units of $(\tilde{\mbox{$m$}}/\mbox
{100 GeV})^2$, $\tilde{m}$ being the mass scale of the scalar
particles with $R_p=-1$.  For the bounds on higher-dimensional
operators we followed Ref.~\cite{blr,bhn}. They state
$\Psi_{112i}~\leq ~10^{-8}~{\tilde{m}}/{\mbox{100 GeV}}$, deriving 
from the experimental bounds on proton
 decay. To take into account the large uncertainties in determining
 this bound (see {\it e.g.} Ref.~\cite{weinberg3}), we worked with 
the
 more conservative
\begin{equation}
\Psi_{112i}~\leq~ 10^{-7}~
\frac{\tilde{m}}{\mbox{100 GeV}}\;.
\end{equation} 
In addition, we always included a factor of tolerance $1/\sqrt{\eps}
\sim 2$ to take into account numerical prefactors $\mcal{O}\!(1)$. 
We initially set ${\tilde m}=100\,{\rm GeV}$ for all bounds and then
indicate whether a model is not ruled out if we allow 100\,GeV$
<\tilde
{m}\leq500\,$ GeV. Heavier sfermion masses lead to weaker bounds.

\item Concerning Step 5, we shall work with the approximation that 
the VEVs of the left-handed sneutrinos vanish as has always been done
in the past (except in Ref.~\cite{bdls}).  This is motivated from
low-energy phenomenology: the coupling constant of the soft
supersymmetry breaking bilinear term with a scalar $L^i$ and a scalar
$H^\mcal{U}$ has to be very small compared to the Higgs-VEV, and the
soft supersymmetry breaking $4\times4$ mass matrix of the scalar
$(H^\mcal{D},L^i)$ has to be nearly diagonal. Hence the VEVs of the
left-handed sneutrinos have to be very small, see Ref.~\cite{gh} and
also Ref.~\cite{add3}. When considering the neutrino masses generated
through the $K_i$ in order to be consistent we must then include the
VEVs for the scalar neutrinos.

\item The  matrices that biunitarily diagonalize the quark mass 
matrices,
namely $\bsym{U^{(U_L)}}$, $\bsym {U^{(U_R)}}$, $\bsym{U^{(D_L)}}$ 
and
$\bsym{U^{(D_R)}}$ are given in Ref.~\cite{hr} as a function of the
entries of the quark mass matrices, as well as $\bsym{U^{C\!K\!M}}=
\bsym {U^{(U_L)}}^\dagger\bsym{U^{(D_L)}}$. We shall not state the 
corresponding expressions here, because they are long and 
complicated,
being a function of the entries of the quark mass matrices. Since we
do not know the $\mcal{O}\!(1)$ coefficients and are hence only
performing order-of-magnitude estimates, we set all coefficients 
equal
to $1$; however, to avoid accidental cancellations we thus have to
replace all minus-signs in the expressions in Ref.~\cite{hr} (this is
also the procedure adopted in previous work). We have checked that 
the
corresponding results are in good agreement with the exact results,
having chosen a randomly generated set of $\mcal{O}\!(1)$ 
coefficients
(see Appendix 2).

\item Due to the uncertainties on masses and mixings in the neutrino
 sector we shall neither use radiative corrections to neutrino masses
 (based on $\not\!\!R_p$ coupling constants) nor Dirac- and
 Majorana-masses (when considering right-handed neutrinos) to 
constrain
 the $X$-charges.
\end{enumerate}

It is important to note that all simplifications which we have 
adopted
here lead to a more conservative approach than calculating exactly
(assuming that there are no accidental cancellations).

Now the various transformations (apart from renormalization group
flow) in Steps~1-7 are all unitary or at least almost unitary. Thus
the determinant of the mass matrices should not change substantially,
so that {apart from a few accidental cancellations and ``accidental
adding-ups''} (and apart from supersymmetric zeroes being filled up)
the entries of the mass matrices (and also the entries of the other
Yukawa coupling constants) \emph{on the average} keep their order of
magnitude. In order to estimate the possible ``accidental 
adding-ups''
in our conservative approach, we have calculated the effects of 
Steps
1-7 on the coupling constants to higher order in $\eps$ (which can
contribute quite a lot as $\epsilon$ is not a very small number). We
used the approximate expressions of the various transformations as 
for
example Eq.~(\ref{bilara}) but kept all powers of $\eps$ in the
result.  Consider {\it e.g.} the charge assignment of Ref.~\cite{blr}
which is displayed at the end of the last section. $\bsym{G^{(U) }}$
is given by Eq.~(\ref{oins}), and $\bsym{C^{(Q) }}^{-1}$, $\bsym{C^{
(\ol{U})}}^{-1}$ are given by ({\it c.f.} Eq.~(\ref{bilara}))
\begin{equation}
\left(\begin{array}{ccc}
1 & \eps & \eps^3 \\
\eps & 1 & \eps^2 \\
\eps^3 & \eps^2 & 1 \end{array}\right),
\left(\begin{array}{ccc}
1 & \eps^3 & \eps^5 \\
\eps^3 & 1 & \eps^2 \\
\eps^5 & \eps^2 & 1 \end{array}\right) ,
\end{equation}
$\bsym{G^{(U)}}$ is then ({\it c.f.} Eq.~(\ref{gupost}))
\begin{eqnarray}
\left(\begin{array}{ccc} 9~\eps^8 & 6~\eps^5+3~\eps^{11} &
 3~\eps^3+3~\eps^7~+3~\eps^{13}\\
6~\eps^7+3~\eps^{9} & 4\eps^4+2\eps^6+2\eps^{10}+
\eps^{12} & 2\eps^2+\eps^4+2\eps^{6}+\eps^{8}+
2\eps^{12}+\eps^{14}\\
3\eps^5+3~\eps^9+3~\eps^{11} & 2\eps^2+2\eps^6+
3\eps^{8}+\!\eps^{12}+\!\eps^{14} & 1\!+2\eps^4\!+\eps^6+
\eps^8+2\eps^{10}\!+\eps^{14}\!+\eps^{16}\end{array}\right).
\nonum
\end{eqnarray}
In the past (and in our conservative approach as well) all numerical
prefactors and all higher orders in $\eps$ were neglected. If 
they are
taken into account one gets with $\eps\sim0.22$ that
\begin{equation}
\label{labelfuer16}
\bsym{G^{(U)}}\sim\left(\begin{array}{rrr}
\eps^{6.5 } &\eps^{3.8 } & \eps^{2.3 } \\
\eps^{5.8 } &\eps^{3.1 } & \eps^{1.5 } \\
\eps^{4.3 } &\eps^{1.5 } & 1 \\
\end{array}\right),
\end{equation}
differing drastically from the conservative result, which is the same
as in Eq.~(\ref{oins}). This is not meant as a demonstration that 
the 
conservative approach produces the wrong mass matrix, it is just an 
indication that \emph{some} of its entries \emph{potentially} can 
get quite large due to ``accidental adding-ups''.

Using the same $X$-charge assignment as before, we give an example
which illustrates how important it is to take into account the
$L^i\lra H^\mcal{D}$ mixing when canonicalizing the K\"ahlerpotential
which has been ignored in the past.  For the $\not\!\!\!R_p$ coupling
constants we obtain
\begin{eqnarray}\label{lammmmda'}
{\Lam^\prime}_{i1k}= 0,~~~{\Lam^\prime}_{i2k}=0,~~~
{\Lam^\prime}_{i3k}\sim\eps^{55}~{G^{(D)}}_{\!ik}\,.
\end{eqnarray}
The model has no right-handed neutrinos so that we do not have to
perform Step~1 which is outlined above. To canonicalize the
K\"ahler\-potential (Step~2) the basis transformation for the
$H^\mcal{D}$ and $L^i$ fields is given by
\begin{eqnarray} 
{\bsym{C^{(L,H^\mcal{D})}}}^{-1}=\left(\begin{array}{llll}
1 &\eps^{12} &\eps^{13} &\eps^{55} \\
 \eps^{12}  & 1 &\eps &\eps^{67} \\
 \eps^{13}  &\eps & 1 &\eps^{68} \\
 \eps^{55} &\eps^{67} &\eps^{68} & 1  \end{array}\right)\,.
\end{eqnarray}
This matrix mixes $\bsym{G^{(D)}}$ with the $\bsym{\Lam^\prime}$
given in Eq.~(\ref{lammmmda'}):
\begin{eqnarray}
{\bsym{C^{(L,H^\mcal{D})}}}^{-1}\cdot\left(\begin{array}{r}
\bsym{G^{(D)}}\\0~~~~~\\0~~~~~\\\eps^{55}~\bsym{G^{(D)}}
\end{array}\right)~=~\left(\begin{array}{l}
~~~~~~\bsym{{G^{(D)}}}+\eps^{10}\bsym{{G^{(D)}}}\\~~
\eps^{12}~\bsym{G^{(D)}}+\eps^{122}\bsym{{G^{(D)}}}  
\\~~ \eps^{13}~\bsym{G^{(D)}}+\eps^{123}\bsym{{G^{(D)}}}
  \\2~\eps^{55}~\bsym{G^{(D)}}\end{array}\right).
\end{eqnarray}
Dropping higher order terms in $\eps$ and neglecting $\mcal{O}\!(1)$
 prefactors this gives
\begin{eqnarray}
\left(\begin{array}{r}
\bsym{{G^{(D)}}}\\\eps^{12}~\bsym{G^{(D)}}  \\ 
\eps^{13}~\bsym{G^{(D)}}  \\\eps^{55}~\bsym{G^{(D)}}
\end{array}\right).\nonumber\end{eqnarray}
With $\eps=0.22$ we conclude that the largest ${\Lam^\prime}_{ijk}$ 
is of 
$\mathcal{O}\!(10^{-8})$. This leads to rapid proton decay, because 
it
turns out that the largest ${\Lam^{\prime\prime}}_{ijk}$ is of 
$\mcal{O}\!
(10^{-7})$. So the model has to be regarded as incompatible with
broken $R$-parity, which is why it is labeled as ``{\it no}'' in the
Table in Appendix~A. Had we not taken into account the $L^i\lra
H^\mcal{D}$ mixing (hence $\bsym{G^{(D)}}$ not mixing with
${\Lam^\prime}_{ijk}$ ) we would have obtained
$$
\left(\begin{array}{r}\bsym{G^{(D)}}
\\\eps^{122}~\bsym{G^{(D)}} \\\eps^{123}~\bsym{G^{(D)}} 
\\\eps^{55}~\bsym{G^{(D)}} \end{array}\right)
$$
instead, giving ${\Lam^\prime}_{ijk}\leq\mcal{O}\!(10^{-38})$, which
is in agreement with current experimental bounds on proton decay. The
next step in the canonicalization of the K\"ahlerpotential would be 
to
replace $\bsym{G^{(D)}}$ by ${\bsym{C^{(Q)}}^{-1}}^T\!\cdot\!
\bsym{G^{(D)}}\!\cdot\!{\bsym{C^{(\ol{D})}}^{-1}}$, 
and likewise ${\Lam^\prime}_{ijk}$, but we shall not go into the 
details 
anymore.

\section{Results}
\setcounter{equation}{0}%
We have taken the existing Froggatt-Nielsen models given in
Refs.~[1-16] and embedded them into our general framework. As
explicitly given in Table~1, after taking into account all the
constraints from the quark masses and mixings, and the lepton masses,
as well as the Green-Schwarz anomaly cancellation conditions, all
$U(1)_X$ charges and thus all models can be expressed in terms of
three integer parameters and three $U(1)_X$ charges which we choose 
to
be ({\it c.f.} footnote~$^{\ref{soertien}}$)
\begin{equation}
x,\,y,\,z \; \in \;\mathsf{Z}\!\!\mathsf{Z},\qquad {\rm and} \qquad 
X_{L^1},\,
X_{L^2},\,X_{L^3}\,.
\label{model}
\end{equation}
The constraints on the integer parameters are given in
Eq.~(\ref{models}), as well as $z=0,1$. The specific choice fixes the
quark mass matrices given in Eqs.~(\ref{oins})-(\ref{fihr}) as well
 as
the charged lepton mass matrix. As a first step, in columns 2-7 of
the  Table~A in Appendix~A we determine the parameters
$x,\,y,\,z$ and the charges $X_{L^i}$ for all models in the
literature. This makes it possible to easily compare them. In
Appendix~A, we give a detailed discussion of the models and how the
charges were reconstructed out of the given information.  In
Table~\ref{catalogue}, we have models with small integer charges ($<$
10) as in model [1], the first model in [4], the models in [5], [7],
[14], and [15]. Models with fractional charges constitute the
rest. Note that $X_{L^1},X_{L^2},X_{L^3}$ being integer is only a
necessary but not sufficient condition for all $X$-charges of
$M\!S\!S\!M$ particles to be integer. In order to see this, it is
important to go back to Table~1 and observe that only a special
combination of $x,y,z$ and the $X_{L^i}$-charges guarantees that
$X_{Q^1}$ and $X_{H^{\mathcal{D}}}$ are integer. This is why {\it
e.g.} the third to last model in Ref.~[3] does not have purely 
integer
$X$-charges, although $X_{L^1}=6,X_{L^2}=1,X_{L^3 }=4$.

Having defined a unified formalism for all models, we developed a
numerical code which implements the steps outlined in
Section~\ref{procedure} for an arbitrary model defined by the choice
in Eq.~(\ref{model}).  We thus determine for each model the order of
magnitude of the $R$-parity violating Yukawa couplings and the
dimension-five operators at the GUT scale. After running the
renormalization group equations one then obtains the corresponding
weak scale predictions.  Of course this occurs under the assumption
that there is no additional discrete symmetry such as $R$-parity or
baryon-parity which forbids a subset of terms.  We can then compare
these predictions with the existing experimental bounds on the
$\rpv$-couplings and the dimension-five operators and thus either
allow or exclude a given model of fermion masses.

For single $R$-parity violating couplings the weak scale bounds have
been converted to GUT scale bounds in Ref.~\cite{add2} using the RGEs
of Ref.~\cite{add}.  We thus compare directly to these. However, we
necessarily predict more than one non-zero $R$-parity violating
coupling at the weak- or the GUT-scale and we must take into account
the bounds on products of couplings.  These can often be much 
stricter
than bounds on single couplings due to extra flavour changing neutral
current effects \cite{add2,dpt,dpt2,Agashe:1995qm} and of course the
strict bounds from proton decay \cite{sv}. In column 8 of
Table~\ref{catalogue} in Appendix~A (``Compatible with Exp.'') we 
list
whether the model satisfies the experimental constraints on baryon-
and lepton-number violation or not.
\begin{itemize}
\item``{\it yes}'' means that the model is compatible with the 
experimental
constraints after performing the procedure outlined in
Sect.~\ref{procedure};\
\item ``{\it (no)}$_{h.o.}$'' means that the model is compatible 
if one 
calculates conservatively, but not compatible if one takes into
account the summation of higher orders in $\eps$ (see the example in
the previous section). A moderate amount of fine-tuning such that
higher order terms cancel would make the model
compatible.\footnote{Strictly speaking taking higher orders into
account also rules out the mass matrices from which one starts out,
see Eq.~(\ref{labelfuer16})}
\item  ``{\it no}'' means that the model is incompatible even to 
lowest 
order in $\eps$.
\item ``{\it yes}/{\it no}'' means that the model is compatible if 
one 
uses the freedom to choose $X_{L^3}$ correspondingly, but the model 
is
not compatible if the particular value of $X_{L^3}$ is used which is
given in the table.
\item The models denoted by $^\ast$
change their status when the relevant sfermion mass is increased.
The details are given in the table caption.
\end{itemize}

\subsection{Discussion}

We find that only three models from Refs.~[1-9] (the models without
right-handed neutrinos) pass with flying colors, {\it i.e.} produce a
``{\it yes}''. {\it (i)} One model of Ref.~\cite{blr} is effectively
$R$-parity conserving due to the very high value of $X_{L^3}$. {\it
(ii)} The model in Ref.~\cite{cch} is compatible due to complicated
fractions for the $X$-charges. {\it (iii)} The model \cite{mnrv} also
survives due to complicated fractional charges. 

Many models, like most of the models in Ref.~[7], are incompatible
solely due to the bounds on higher dimensional operators. As an
example, consider the first model of Ref~[7]: $x=y=z=0,\, X_{L^1}=
X_{L^3}=-6,$ and $X_{L^2}=-3\;$ give $X_{Q^1}=5$, $X_{Q^2}=4$, so 
that
the operator $Q^1Q^1Q^2L^3$ is suppressed by $\epsilon^8$, which is a
factor $\sim 55$ too large.

It is important to note that including the exact transformation 
matrix
to the canonical basis for the kinetic terms, $\bsym{C}^{...}$, 
as well as the exact matrix
for rotating away the $LH^\mcal{U}$-term makes it even harder for
models to survive.

In the case without right-handed neutrinos, we see that a flavour
dependent $U(1)_X$ cannot solve the problem of baryon- and
lepton-number violation if we demand small integer or simple
fractional $X$-charges. We have thus also considered the case of an
additional discrete symmetry: $R$-parity, baryon-parity, or
lepton-parity \cite{ir}. In the last column of Table~\ref{catalogue},
we show which set of couplings must vanish for the model to
survive. Thus for example the second set of models in Ref.~[1] 
survive
(for certain n) if $\Psi=\Lam''=0$, which can be guaranteed by
baryon-parity. Similarly, the third model in Ref.~[3] is compatible
with the experimental constraints if $\Lam=\Lam'=0$, which can be
achieved by lepton-parity (anomalous). The first model in Ref.~[7] becomes
compatible if $\Psi=\Lam'=0$, which is also obtained by
lepton-parity. In this manner, we can obtain models with small 
integer
charges, but we also lose a central part of our motivation.  If there
is no entry in the last column the model is either compatible or
beyond repair.

\medskip

When we allow for right-handed neutrinos, Refs.~[10-16], we find a
significant number of allowed models. Here the $X$-charges of the
right-handed neutrinos present extra degrees of freedom, making the
charge assignments more flexible to avoid bounds on the $\not\!\!\!R
_p$ coupling constants. (If one does not have right-handed neutrinos
one is forced to have non-vanishing $\Lam_{ijk}$ and/or
${\Lam^\prime}_{ijk}$ in order to get neutrino masses.) However, again
we find that the models with integer $X$-charges ([14] and [15]) are
not compatible.  The models [11,12] automatically have conserved
$R$-parity and no $QQQL$ operator, {\it i.e.} $\Psi=0$, due to the
fractional charges (although $X_{L^1},X_{L^2},X_{L^3}$ are integers).
On the other hand, while the models in [10,2] certainly can fulfill
all constraints if one chooses an appropriate $X_{L^3}$, certain
choices of $X_{L^3}$ providing special realistic cases are in 
conflict
with the constraints. The charges are not ``fractional enough'', that
is, they still produce positive-integer exponents of $\epsilon$ for
certain $\not\!\!R_p$ coupling constants, thus not forbidding
$\not\!\!R_p$ by fractional $X$-charges as is the case in 
Refs.~[11,12].

For completeness in Table~B  we also investigated
and listed  the  Froggatt-Nielsen  models  (one in Ref.~[16],
the rest  in   Refs.~[17-21])   which  exhibit  a slightly 
different fermionic  mass  spectrum  than the  models in 
Refs.~[1-16].

%
%
\section{Summary}
We have presented a systematic embedding of four-dimensional 
Froggatt-Nielsen models
with a single Froggatt-Nielsen field into supersymmetry. We have
developed a simple notation into which we have translated all
models  [fulfilling Eqs.~(1.1-1.7) and the Green-Schwarz conditions]
which we found in the literature. We have then extended 
(where necessary) the
models to include their predictions for baryon- and lepton-number
violation, including the dimension-five operators. Throughout, we 
have
taken into account the modification of the fermion mass matrices due
to canonicalization of the K\"ahler potential, and the rotation of 
the
$H^{\cal D}$ and $L_i$ fields. We also consider the case of summing
sub-leading powers of $\eps$, which can lead to significant
effects. We have outlined in detail how the models are translated 
from
high-energy textures to low-energy predictions.  We have discussed 
the
required steps and pointed out where we have improved on the existing
literature.  In our Table in Appendix~A, we show how many of the
existing models fair when compared to the most recent bounds on $B$-
and $L$-parity violating coupling constants.  Only very few models
survive unscathed. In the future we hope to improve on our procedure
as mentioned in the steps.  We also wish to extend our discussion to
neutrino masses which we have not yet taken into account.
%
%
%
%
\section{Acknowledgments}
\setcounter{equation}{0}
A special vote of thanks goes to Martin Walter for helpful 
discussions.
M.T. would like to thank the Evangelisches  Studienwerk and 
Worcester 
College for financial support during his time  at the 
Subdepartment of Theoretical Physics of the University of Oxford, 
while 
part of this work was completed.
%
%
%

\begin{appendix}

\section{\label{catalogue}The Catalogue of Froggatt-Nielsen Models}
The first nine references deal with models with no right-handed
neutrinos, so no $\Xi_i~{\ol N^i}$ term has to be shifted away. The
values for the six parameters ($x,\,y,\,z,\,X_{L^i}$) were obtained 
as
follows (the references are listed chronologically according to their
appearance on the arXive):

\begin{itemize}
\item \cite{dps} explicitly states four sets of $X$-charges, three 
of these sets fit the class of model considered in this paper (in the
third model there is a typo in the $X$-charge assignments: instead of
$e_1=-1+q,~e_2=5+q$ it should read $e_1=5+q,~e_2=-1+q$). The
construction of these models did not take into account the
Giudice-Masiero mechanism, higher-dimensional operators and $\not\!\!
R_p$. It turns out that the models are in conflict with the bounds on
$\Psi$ independent of the value chosen for $X_{L^3}$. For low values
of $|X_{L^3}|$, the models also disagree with bounds on 
$\not\!\!R_p$.

\item \cite{blr} states two complete sets of $X$-charges. The 
authors 
take into account higher-dimensional and $\not\!\!\!R_p$-operators as
well as the Giudice-Masiero mechanism. The second model is in accord
with all constraints due to the large value of $|X_{L^3}|$. The first
model was treated as an example in Section~4. The paper presents a
further model with right-handed neutrinos that is also discussed in
Ref.~\cite{l1}, see below.

\item \cite{cl} considers the choice $y=-6,0$ (in our notation). The 
corresponding quark mass matrices must be combined with four 
different
explicitly stated charged lepton mass matrices. These are obtained
through four different choices for the charge differences $X_{L^1}-X_
{L^3}$ and $X_{L^2}-X_{L^3}$. The first model is given (in our
notation) by $x=0$, the second one has $y=-6$ with $x=0 ,1,2$. The
third model is required to have $x=0$, and for the last one $x=0,$ and
$y=-6$ is demanded. In addition, the authors work with $\ell_0\equiv
X_{L^3}+ X_{Q^3}+X_{{\ol D}^3}$ to be $4$ or $5$, $-4$ or $-3$, $7$ or
$8$, and $-7$ or $-6$, respectively. They thus have $2\cdot(2+3+2+1)
=16$ $X $-charge assignments. All are such that no dangerous
higher-dimensional operators arise. But the authors do not take into
account the Giudice-Masiero mechanism, the $LH^\mcal{U}$-term, or the
mixing of $L^i$ and $H^\mcal{D}$. They do consider $\not\!\!R_p$
(explicitly needed to generate neutrino masses), however twelve 
models
disagree with the experimental bounds at lowest order in $\eps$ and
the four other models disagree when including higher order terms.

\item \cite{cck} explicitly states the three sets of $X$-charges.
 The authors take into account higher-dimensional operators as well 
as
 $L^i\lra H^\mcal{D}$ mixing and thus mixing of $QQQL$ and 
$QQQH^\mcal
 {D}$. The first two models are in accord with the bounds to lowest
 order in $\eps$, however not so to higher order.  The third model is
 ruled out because it does not agree with $\not\!\!R_p$: $X_{L^1}+X_
 {L^3}+X_{E^1}=0$ and thus $\Lam_{131}={\cal O}(1)$.

\item \cite{bdls} explicitly states an incomplete but sufficient
 set of $X$-charges. They disregard higher-dimensional operators, and
 the corresponding constraints are not met.
  
\item \cite{cch} explicitly states two sets of $X$-charge assignments
 which however are not compatible with the Green-Schwarz
 mechanism. But from their requirements (translated to our notation)
 of $y=0,\, z=1$, $X_{L^1}-2=X_{L^2}=X_{L^3 }\,$ and $7\leq X_{L^3}-
 X_{H^\mcal{D}}\leq 9$ and $\tan\beta\approx 50~\Rightarrow~x
 = 0$ one can extract one $X$-charge assignment which is
 compatible with the bounds.
  
\item \cite{jvv} states 66 (!) $X$-charge assignments, many of them 
 however referring to quark mass matrices that deviate slightly from
 the ones considered here: The naive exponents in $\bsym{G^{(U)}}$ 
and
 $\bsym{G^{(D)}}$ are given as follows
\begin{eqnarray}
\left(\begin{array}{rrr}
8 & \phantom{-}5 & \phantom{-}4 \\
7 & 4 & 3  \\
4 & 1 & 0 \end{array}\right), & &~~~\left(\begin{array}{rrr}
5 & {3} & \phantom{-}{4} \\
4 & 2 & 3 \\
1 & -1 & 0 \end{array}\right),\nonum\\
\left(\begin{array}{rrr}
8 & \phantom{-}5 & \phantom{-}4 \\
7 & 4 & 3  \\
4 & 1 & 0 \end{array}\right),& &~~~\left(\begin{array}{rrr}
3 & {3} & {4} \\
2 & 2 & 3 \\
-1 & -1 & \phantom{-}0 \end{array}\right),\nonum\\
\left(\begin{array}{rrr}
8 & \phantom{-}5 & -2 \\
7 & 4 & -3  \\
10 & 7 & 0 \end{array}\right),& &~~~\left(\begin{array}{rrr}
4 & \phantom{-}{3} & {-2} \\
3 & 2 & -3 \\
6 & 5 & 0 \end{array}\right).\nonum
\end{eqnarray}
 We shall not consider these charge assignments. Focusing on
 Eq.~(\ref{oins}), we are left with 26 $X$-charge assignments. Note
 that their $2^{nd}$, $12^{th}$ and $15^{th}$ model are special cases
 of the $1^{st}$, $2^{nd}$ and $2^{nd}$ model of Ref.~\cite{dps},
 respectively. Their $8^{th}$ model equals the one in Ref.~[5]. The
 authors did not take into account higher-dimensional operators; the
 corresponding bounds rule out many models. Although the authors do
 consider $\not\!\!R_p$, some of the models are ruled out due to
 disagreement with more recent bounds on $\not\!\!R_p$ coupling
 constants.  In the end, none of the models survives even to lowest
 order.

\item \cite{mnrv} explicitly states a set of $X$-charges. All 
 constraints are carefully obeyed, leading to a valid model. 
The price
 to be paid are highly fractional charges.

\item \cite{r} explicitly states a set of $X$-charges.
The author does not take into account higher-dimensional operators, 
as
the corresponding bounds rule out the model.

\item \cite{l1} and \cite{blr} (the first reference is a talk 
discussing
 one of the models -- this time with right-handed neutrinos -- of the
 second reference) demand $X_{L^1}-3=X_{L^2}-1=X_{L^3}$. This still
 leaves $X_{{L^3}}$ undetermined. Combining this with the fact that
 the authors work with $y=-6,0$, we get with $z=0,1$ that there are
 $2\times(3+4)=14$ charge assignments with $X_{L^3}$ being a free
 parameter. Although explicitly including right-handed neutrinos, no
 corresponding $X$-charges are stated. Of course one might argue that
 it is always possible to choose a value for $X_{{L^3}}$ such that
 $R$-parity is exactly conserved or at least broken in a way that
 experimental limits are obeyed. However, to consider a special
 realistic case, the authors demand the Dirac and Majorana neutrino
 mass matrices to be without supersymmetric zeroes and the
 (3,3)-component to dominate, {\it i.e.} one needs $X_{{\ol{N^i}}}+X_
 {H^ \mcal{U}}+X_{L^j}\geq X_{{\ol{N^3}}}+X_{H^\mcal{U}}+X_{L^3}\geq
 0$ and $X_{{\ol{N^i}}}+X_{{\ol{N^j}}}\geq X_{{\ol{N^3}}}+
X_{{\ol{N^3}
 }}\geq 0$. From Eq.~(14) of Ref.~\cite{l1}, we thus find
\begin{eqnarray}\label{zem}
m_{\nu_\tau}\sim\frac{\big\langle {{{H}^\mcal{U}}}
\big\rangle^2 
~\eps^{2(X_{{H^\mcal{U}}}~+~
X_{L^3})}}{M}\;.
\end{eqnarray}
 With $y=0,-6$ the $X$-charge assignments are not GUT compatible, one
 thus assumes $M=M_{\!s}\sim10^{18}$ GeV (and not 
$M=M_{\rm GUT}\sim
 10^{16}$ GeV). The authors demand $m_{\nu_\tau}=0.1$ eV. Combining
 this with $\eps\sim 0.22, ~\big\langle {H}^\mcal{U}\big\rangle\sim10
 0$ GeV, we get that $X_{H^\mcal{U}}+X_{L^3}\sim-3$; we can thus fix
 $X_{{L^3}}$.
  
\item \cite{l2} and \cite{blpr} (the first reference is a talk 
discussing 
 one of the models of the second reference) present a model (aimed at
 atmospheric neutrinos) that is based on the quark sector in
 Ref.~\cite{blr} (Ref.~\cite{blpr} is a kind of continuation of
 Ref.~\cite{blr}.  Ref.~\cite{l2} also reviews the model of
 Refs.~\cite{blr,l1}), so $y=-6,0$.  The authors demand $\frac{3}{2}
 \leq X_{L^2}=-X_{L^3}\leq\frac{5}{2},\,X_{L^1}-X_{L^2}\geq3$, which 
 they specify to $X_{L^1}= 5, X_{L^2}=2$, giving $X_{{\ol{E^1}}}-X_{ 
 {\ol{E^2}}}=z-1$. As they furthermore suppose $X_{{\ol{E^1}}}=
X_{{\ol
 {E^2}}}$, the authors get $z=1$. This leads to $m_e/m_\tau\sim\eps^5
 $, as is correctly stated in Ref.~\cite{blpr}. It should however be
 mentioned that both Refs.~\cite{l2} and \cite{blpr} explicitly 
demand
 $X_{H^\mcal{U}}=X _{H^\mcal{D}}=0$, which is in contradiction to
 $z=1$.  Furthermore, in Ref.~\cite{blpr} it is assumed that $2\leq
 X_{{\ol{E^3}}}\leq 4$, which requires $0<143+29x+3x^2-18y<120+18x$,
 which is not possible.  Therefore the model is actually
 self-contradicting. However, Ref.~\cite{l2} has not stated this
 constraint on $X_{{\ol{E^3}}}$ and it shall hence be ignored. This
 gives 4+3=7 different models. The constraints on the $X$-charges of
 the right-handed neutrinos are $0\leq X_{{\ol{N^1}}}<-X_{{\ol{N^3}}}
 \leq X_{L ^2}\leq X_{{\ol{N^2}}}$.
  
\item \cite{blpr} furthermore considers a model that was not 
 dealt with in Ref.~[11], aimed at the adiabatic
 Mikheev-Smirnow-Wolfenstein effect \cite{msw1,msw2} for the solar
 neutrinos. The authors demand $0<X_{L^2}<-X_{L^3}= X_{L^1}$ and the
 following constraints on the $X$-charges of the right-handed
 neutrinos: $0\leq X_{\ol{N^1}},~X_{L ^2} <- X_{\ol{N^3}}\leq
 X_{L^1}\leq X_{\ol{N^2}}$. Specifically they use
 $X_{L^1}=\frac{9}{2},X_{L^2}=\frac{3}{2},X_{\ol{E^1}}=X_{\ol{E^3
 }}-4$. This results in $z=1$ and $X_{\ol{E^1}}=X_{\ol{E^2}}\,$;
 furthermore the authors demand $\frac{9}{2}\leq X_{\ol{E^3}} \leq
 \frac{17}{2}$, which is fulfilled. So there are 4+3=7 models.
  
\item \cite{eir} indirectly  states that $x=3,\,y=0,\,z=0$ and gives
  an explicit charged lepton mass matrix and a right-handed neutrino
  Majorana mass matrix. The latter depends on a parameter labeled
  $X^{[\bar{N}]}$, which is given as $-\frac{9}{2}$, in order to
  accomodate the non-adiabatic MSW effect. This completely fixes the 
  $X_{L^i}$, and one has  $X_{\ol{N^1}}=\frac{13}{2},X_{\ol{N^2}}
  =\frac{11}{2},X_{\ol{N^3}}=\frac{3}{2}$.
  
\item \cite{hm} states an explicit $X$-charge assignment, with
  right-handed neutrino charges $X_{\ol{N^1}}=X_{\ol{N^2}}=X_{\ol{N
  ^3}}=0$. Note that totally uncharged fields may give rise to
  dangerous tadpoles, {\it e.g.} see Refs.~\cite{tp1,tp2,tp3,tp4}.
  
\item \cite{bs} states an explicit $X$-charge assignment, with
  right-handed neutrino charges $X_{{\ol{N^1}}}=8,X_{{\ol{N^2}}}=4,X_
  {{\ol{N^3}}}=0$. These $X$-charges are the only difference to the
  model in Ref.~\cite{hm}. This model and the previous one are the
  only ones that have the quark mass matrices of Eq.~(\ref{droi}).

\item \cite{ky} explicitly demands conserved $R_p$, however without  
 checking whether this is a consequence of the given $X$-charge
 assignments. The authors also do not constrain the $X$-charges by
 anomaly cancellation, which gives them more freedom. We reconsider
 their model taking the Green-Schwarz mechanism into account. The
 authors indirectly state that $z=1,y=0$ and give several $X$-charges
 parameterized by the integers $m,n,p,r$. Then these parameters are
 fixed to be $p=r=5,~n=0$ (so-called ``best-fit assignment'').
 One then distinguishes  $m=
 0$ or $1$ (``anarchical type'' or ``lopsided type'', respectively).
 $x$ and $X_{L^3}$ are left undetermined. The authors thus have $2
 \times 4=8$ models with $X_{L^3}$ unconstrained, so one should 
 always be able to choose $X_{L^3}$ such that $R_p$ is conserved. In
 the end of the paper the authors explicitly emphasize that $\tan 
 \beta\sim3~\Rightarrow~x=3$, together with $2X_{Q^3}-X_{{\ol{U^3}}}
- 
 X_{{\ol{E^3}}}\stackrel{!}{=}1$. This seems to give a way to fix 
$X_{L^3}$; however with Table~1 one gets $2X_{Q^3}-X_{\ol{U^3}}-
X_{\ol{E^3}}
=\frac{4-m}{3}$, independent of $X_{L^3}$ (note that one actually 
needs to 
have  $m=1$,
otherwise the model is inconsistent). So we had to leave  $X_{L^3}$
unfixed. Although explicitly including
 right-handed neutrinos, no corresponding $X$-charges are stated.
\end{itemize}
\begin{center}
\begin{tabular}{|c|c|c|c|c|c|c|c|c|}
\hline
$\phantom{\Bigg|}$ {\bf{by}} $\phantom{\Bigg|}$  &            
         ${\bsym{x}}$                            &                  
         ${\bsym{y}}$                            &                   
         ${\bsym{z}}$                            &                  
         ${\bsym{X_{L^1}}}$                      &
         ${\bsym{X_{L^2}}}$                      &
         ${\bsym{X_{L^3}}}$                      & 
          {
           \begin{tabular}{c} 
             {\bf Compatible}         \\
             {\bf with Exp.} \end{tabular} }     &
          {\bf O.K. if ...} \\
\hline
$\phantom{\Bigg|}$\cite{dps}$\phantom{\Bigg|}$ & $ 0 $ & $0$ & $0$ & 
$X_{L^3}$ & $X_{L^3}$ & \begin{tabular}{c}$1-3$n,
\\n$\in\mathsf{Z}\!\!\mathsf{Z}$\end{tabular} & 
~~{\it no}\phantom{$^*$}~ 
if~ 
n$\neq-3,\ldots,1~~$ 
 & $\Psi=0$ \\
$\phantom{\Bigg|}$ $\phantom{\Bigg|}$ & $  $ &  &  &  &  &  & 
~~{\it no}\phantom{$^*$}~ 
if~  n$=-3,-2,-1$  & 
$\Lam^{\prime\prime},\Psi=0$ \\
$\phantom{\Bigg|}$ $\phantom{\Bigg|}$ &  & & &  &  &  & 
~~{\it no}\phantom{$^*$}~ if~ 
n$=0,1~~~~~~~~~~$  & \\
\hline
$\phantom{\Bigg|}$ $\phantom{\Bigg|}$ & $ 2 $ & $0$ & $0$ & 
$X_{L^3}+1$ & $X_{L^3}-1$ & \begin{tabular}{c}$3-3$n,\\
n$\in\mathsf{Z}\!\!\mathsf{Z}$\end{tabular} &
{\it no}$^*$~ if~ n$\neq-2,\ldots,2$ 
 & $\Psi=0$ \\
$\phantom{\Bigg|}$ $\phantom{\Bigg|}$ & $  $ &  &  & 
 &  &  &
~~{\it no}\phantom{$^*$}\phantom{$^*$}~ if~ n$=-2,-1~~~~$  & 
$\Lam^{\prime\prime},\Psi=0$ \\
$\phantom{\Bigg|}$ $\phantom{\Bigg|}$ & $  $ &  &  & 
 &  & &
~~{\it no}\phantom{$^*$}\phantom{$^*$}~ if~ n$=0,1,2~~~~~~$  & \\
\hline
$\phantom{\Bigg|}$ $\phantom{\Bigg|}$ & 2 & 0 & 0 & 
$X_{L^3}+1$ &  $X_{L^3}+5$ &\begin{tabular}{c}$1-3$n,\\
n$\in\mathsf{Z}\!\!\mathsf{Z}$\end{tabular} &
{\it no}\phantom{$^*$}~ if~ n$\neq-2,\ldots,2$ 
 & $\Psi=0$ \\
$\phantom{\Bigg|}$ $\phantom{\Bigg|}$ &  &  & & 
 &  & &
{\it no}\phantom{$^*$}~ if~ n$=-2,-1~~~$  & 
$\Lam^{\prime\prime},\Psi=0$ \\
$\phantom{\Bigg|}$ $\phantom{\Bigg|}$ &  &  &  &  &  &  &
{\it no}\phantom{$^*$}~ 
if~ n$=0,1,2~~~~~$  & \\
\hline
$\phantom{\Bigg|}$\cite{blr}$\phantom{\Bigg|}$ & $2$ & $0$ & $0$ & 
$-12\phantom{-}$ 
& $-13\phantom{-}$ &  $55$  & {\it no} & $\Lam^{\prime\prime}=0$\\
\hline 
$\phantom{\Bigg|}$$\phantom{\Bigg|}$ & $2$ & $0$ & $0$ & $23$ & 
$22$ &  $-78\phantom{-}$ & {\it yes}  & \\
\hline
$\phantom{\Bigg|}$\cite{cl}$\phantom{\Bigg|}$ & $0$ & $0$ & $0$ & 
$\frac{32}{5}$ & $\frac{22}{5}$ &  $\frac{7}{5}$  &
 {\phantom{$_{h.o.}$}\it (no)$_{h.o.}$}  & $\Lam^\prime=0$ \\
\hline
$\phantom{\Bigg|}$$\phantom{\Bigg|}$ & 0 & $-6\phantom{-}$ & $0$ & 
$\frac{26}{5}$ 
& $\frac{16}{5}$ &  $\frac{1}{5}$  & {\it no} & $\Lam^\prime=0$ \\
\hline
\end{tabular}$~$\\
\emph{continued next page}
\begin{tabular}{|c|c|c|c|c|c|c|c|c|}
\hline
$\phantom{\Bigg|}${\bf{by}}$\phantom{\Bigg|}$&                     
         ${\bsym{x}}$                           &                  
         ${\bsym{y}}$                           &                  
         ${\bsym{z}}$                           &                  
         ${\bsym{X_{L^1}}}$                     &
         ${\bsym{X_{L^2}}}$                     &
         ${\bsym{X_{L^3}}}$                     & 
          {
           \begin{tabular}{c} 
             {\bf Compatible}         \\
             {\bf with Exp.} \end{tabular} }                   &
          {\bf O.K. if ...} \\
\hline
$\phantom{\Bigg|}$$\phantom{\Bigg|}$ & 0 & $-6\phantom{-}$ & $0$ & 
$\frac{62}{5}
$ &  $-\frac{48}{5}\phantom{-}$ &  $-\frac{33}{5}\phantom{-}$ & 
{\it no} 
 & $\Lam,\Lam^\prime=0$ \\
\hline
$\phantom{\Bigg|}$$\phantom{\Bigg|}$ & 1 & $-6\phantom{-}$ & $0$ & 
$\frac{1321}{105}$ 
& $-\frac{989}{105}\phantom{-}$ &  $-\frac{674}{105}\phantom{-}$ & 
{\it no} 
& $\Lam,\Lam^\prime=0$ \\\hline
$\phantom{\Bigg|}$$\phantom{\Bigg|}$ & 2 & $-6\phantom{-}$ & $0$ & 
$\frac{38}{3}$ & 
$-\frac{28}{3}\phantom{-}$ &  $-\frac{19}{3}\phantom{-}$  & {\it no} 
&
 $\Lam,\Lam^\prime=0$ \\
\hline
$\phantom{\Bigg|}$$\phantom{\Bigg|}$ & 0 & $0$ & $0$ & 
$\frac{33}{5}$ 
& $\frac{8}{5}$ &  $\frac{23}{5}$ &  {\phantom{$_{h.o.}$}
\it (no)}$_{h.o.}$
&  $\Lam^\prime=0$ \\
\hline
$\phantom{\Bigg|}$$\phantom{\Bigg|}$ & 0 & $-6\phantom{-}$ & $0$ & 
$\frac{27}{5}
$ & $\frac{2}{5}$ &  $\frac{17}{5}$ & {\it no}  & $\Lam^\prime=0$ \\
\hline
$\phantom{\Bigg|}$$\phantom{\Bigg|}$ & 0 & $-6\phantom{-}$ & $0$ & 
$\frac{61}{5}
$ & $-\frac{34}{5}\phantom{-}$ &  $-\frac{49}{5}\phantom{-}$  & 
{\it no}
 & $\Lam^\prime=0$ \\
\hline
$\phantom{\Bigg|}$$\phantom{\Bigg|}$ & 0 & $0$ & $0$ & 
$\frac{29}{5}$ &
 $\frac{19}{5}$ &  $\frac{4}{5}$ &  {\phantom{$_{h.o.}$}
\it (no)}$_{h.o.}$
 & $\Lam^\prime=0$  \\
\hline
$\phantom{\Bigg|}$$\phantom{\Bigg|}$ & 0 & $-6\phantom{-}$ 
& $0$ & 
$\frac{23}{5}$ & 
$\frac{13}{5}$ &  $-\frac{2}{5}\phantom{-}$ & {\it no} &  
$\Lam^\prime=0$ \\
\hline
$\phantom{\Bigg|}$$\phantom{\Bigg|}$ & 0 & $-6\phantom{-}$ & $0$ & 
$\frac{59}{5}$ & 
 $-\frac{51}{5}\phantom{-}$ &  $-\frac{36}{5}\phantom{-}$  & {\it no}
 & $\Lam,\Lam^\prime=0$ \\
\hline
$\phantom{\Bigg|}$$\phantom{\Bigg|}$ & 1 & $-6\phantom{-}$ & $0$ & 
$\frac{1258}{105}$
 & $-\frac{1052}{105}\phantom{-}$ &  $-\frac{737}{105}\phantom{-}$ 
& {\it no}
 & $\Lam,\Lam^\prime=0$ \\
\hline
$\phantom{\Bigg|}$$\phantom{\Bigg|}$ & 2 & $-6\phantom{-}$ & $0$ & 
$\frac{181}{15}$ & 
$-\frac{149}{15}\phantom{-}$ &  $-\frac{104}{15}\phantom{-}$ & 
{\it no} & 
$\Lam,\Lam^\prime=0$ \\
\hline
$\phantom{\Bigg|}$$\phantom{\Bigg|}$ & 0 & $0$ & $0$ & $6$ 
& $1$ &  $4$ & {\phantom{$_{h.o.}$}\it (no)}$_{h.o.}$  & $\Lam,
\Lam^\prime=0$
 \\
\hline
$\phantom{\Bigg|}$$\phantom{\Bigg|}$ & 0 & $-6\phantom{-}$ & $0$ & 
$\frac{24}{5}$
 & $-\frac{1}{5}\phantom{-}$ &  $\frac{14}{5}$ & {\it no} & 
$\Lam^\prime=0$
 \\
\hline
\end{tabular}$~$\\
\emph{continued next page}
\begin{tabular}{|c|c|c|c|c|c|c|c|c|}
\hline
$\phantom{\Bigg|}${\bf{by}}$\phantom{\Bigg|}$&                     
         ${\bsym{x}}$                           &                  
         ${\bsym{y}}$                           &                  
         ${\bsym{z}}$                           &                  
         ${\bsym{X_{L^1}}}$                     &
         ${\bsym{X_{L^2}}}$                     &
         ${\bsym{X_{L^3}}}$                     & 
          {
           \begin{tabular}{c} 
             {\bf Compatible}         \\
             {\bf with Exp.} \end{tabular} }                   &
          {\bf O.K. if ...} \\
\hline
$\phantom{\Bigg|}$$\phantom{\Bigg|}$ & 0 & $-6\phantom{-}$ & $0$ & 
$\frac{58}{5}
$ & $-\frac{37}{5}\phantom{-}$ &  $-\frac{52}{5}\phantom{-}$ & 
{\it no} & 
$\Lam,\Lam^\prime=0$ \\
\hline
$\phantom{\Bigg|}$\cite{cck}$\phantom{\Bigg|}$ & $3$ & $0$ & $1$ &
 $-8\phantom{-}$ & $-8\phantom{-}$ & 
 $-4\phantom{-}$ & {\phantom{$_{h.o.}$}\it (no)}$_{h.o.}$  & 
$\Psi=0$ \\
\hline
$\phantom{\Bigg|}$$\phantom{\Bigg|}$& $0$ & $0$ & $1$ & 
$-4\phantom{-}$ & 
$-7\phantom{-}$ &  
 $-\frac{9}{2}\phantom{-}$ & {\phantom{$_{h.o.}$}\it 
(no)}$_{h.o.}$  & 
$\Psi=0$  \\
\hline
$\phantom{\Bigg|}$$\phantom{\Bigg|}$& $3$ & $0$ & $1$ & 
$-\frac{7}{2}\phantom{-}$
 &$-4\phantom{-}$& 
$-\frac{7}{2}\phantom{-}$  & {\it no}  & $\Lam=0$   \\
\hline
$\phantom{\Bigg|}$\cite{bdls}$\phantom{\Bigg|}$ & $0$ & $0$ & $0$ & 
$-8\phantom{-}$ 
& $-8\phantom{-}$ &  $-8\phantom{-}$ & {\it no} &$\Psi=0$  \\
\hline
$\phantom{\Bigg|}$\cite{cch}$
\phantom{\Bigg|}$& $0$ & $0$ & $1$ & $ X_{L^3}+2$ & $X_{L^3}$ &  
$\frac{358}{105}\leq ... \leq\frac{478}{105}   $ & {\it yes} & \\
\hline
$\phantom{\Bigg|}$\cite{jvv}$\phantom{\Bigg|}$ & $0$ & $0$ & $0$ & 
$-6\phantom{-}$ & 
$-3\phantom{-}$ &  $-6\phantom{-}$ & {\it no} &
 $\Lam^\prime,\Psi=0$   \\ 
\hline
$\phantom{\Bigg|}$$\phantom{\Bigg|}$ & $0$ & $0$ & $0$ & $-
5\phantom{-}$ & $-5\phantom{-}$ &  $-5\phantom{-}$ & 
{\it no} &$
\Psi=0$ \\
\hline
$\phantom{\Bigg|}$$\phantom{\Bigg|}$ & $0$ & $0$ & $0$ & 
$-4\phantom{-}$ & $-7\phantom{-}$ &  $-4\phantom{-}$& 
{\it no} &$
\Psi=0$\\
\hline
$\phantom{\Bigg|}$$\phantom{\Bigg|}$ & $0$ & $0$ & $0$ & 
$-3\phantom{-}$ & $-9\phantom{-}$ &  $-3\phantom{-}$& 
{\it no} &  $\Lam,\Lam^\prime,\Psi=0$ \\
\hline
$\phantom{\Bigg|}$$\phantom{\Bigg|}$ & $0$ & $0$ & $0$ & 
$-10\phantom{-}$ & $-4\phantom{-}$ &  $-1\phantom{-}$&
 {\it no} &   $\Lam^\prime,\Psi=0$  \\
\hline
$\phantom{\Bigg|}$$\phantom{\Bigg|}$ & $0$ & $0$ & $0$ & 
$-10\phantom{-}$ & $-4\phantom{-}$ &  $-10\phantom{-}$ &
 {\it no}&
 $
\Psi=0$  \\
\hline
$\phantom{\Bigg|}$$\phantom{\Bigg|}$ & $0$ & $0$ & $0$ & 
$-9\phantom{-}$ & $-6\phantom{-}$ &  $-9\phantom{-}$ &
 {\it no}  &$
\Psi=0$ \\
\hline
\end{tabular}$~$\\\emph{continued next page}
\begin{tabular}{|c|c|c|c|c|c|c|c|c|}
\hline
$\phantom{\Bigg|}${\bf{by}}$\phantom{\Bigg|}$&                     
         ${\bsym{x}}$                           &                  
         ${\bsym{y}}$                           &                  
         ${\bsym{z}}$                           &                  
         ${\bsym{X_{L^1}}}$                     &
         ${\bsym{X_{L^2}}}$                     &
         ${\bsym{X_{L^3}}}$                     & 
          {
           \begin{tabular}{c} 
             {\bf Compatible}         \\
             {\bf with Exp.} \end{tabular} }                   &
          {\bf O.K. if ...} \\
\hline
$\phantom{\Bigg|}$$\phantom{\Bigg|}$ & $0$ & $0$ & $0$ & 
$-8\phantom{-}$ & $-8\phantom{-}$ &  $-8\phantom{-}$ &
 {\it no} &$
\Psi=0$ \\
\hline
$\phantom{\Bigg|}$$\phantom{\Bigg|}$ & $0$ & $0$ & $0$ & 
$-7\phantom{-}$ & $-10\phantom{-}$ &  $-7\phantom{-}$  &
 {\it no} &$
\Psi=0$ \\
\hline
$\phantom{\Bigg|}$$\phantom{\Bigg|}$ & $2$ & $0$ & $0$ & 
$-7\phantom{-}$ & $-3\phantom{-}$ &  $-8\phantom{-}$ & 
{\it no}  &$
\Psi=0$  \\
\hline 
$\phantom{\Bigg|}$$\phantom{\Bigg|}$ & $2$ & $0$ & $0$ & 
$-6\phantom{-}$ & $-5\phantom{-}$ &  $-7\phantom{-}$ &
\phantom{$^{\ast}$}{\it no}$^{\ast}$
 &$\Psi=0$ 
\\
\hline
$\phantom{\Bigg|}$$\phantom{\Bigg|}$ & $2$ & $0$ & $0$ & 
$-5\phantom{-}$ & $-7\phantom{-}$ &  $-6\phantom{-}$ &
 \phantom{$^{\ast}$}{\it no}$^{\ast}$  &  
$\Psi=0$  \\
\hline
$\phantom{\Bigg|}$$\phantom{\Bigg|}$ & $2$ & $0$ & $0$ & 
$-4\phantom{-}$ & $-9\phantom{-}$ &  $-5\phantom{-}$ & 
{\it no}  &$\Psi=0$  \\
\hline
$\phantom{\Bigg|}$$\phantom{\Bigg|}$ & $2$ & $0$ & $0$ & 
$-9\phantom{-}$ & $-8\phantom{-}$ &  $-10\phantom{-}$  &
\phantom{$^{\ast}$}{\it no}$^{\ast}$ &
$\Psi=0$  \\
\hline 
$\phantom{\Bigg|}$$\phantom{\Bigg|}$ & $2$ & $0$ & $0$ & 
$-8\phantom{-}$ & $-10\phantom{-}$ &  $-9\phantom{-}$ &
\phantom{$^{\ast}$}{\it no}$^{\ast}$  &
$\Psi=0$  \\
\hline
$\phantom{\Bigg|}$$\phantom{\Bigg|}$ & $0$ & $0$ & $1$ & 
$-6\phantom{-}$ & $-3\phantom{-}$ &  $-2\phantom{-}$& 
 {\it no} & $\Lam,\Lam^\prime,\Psi=0$  \\
\hline
$\phantom{\Bigg|}$$\phantom{\Bigg|}$ & $0$ & $0$ & $1$ & 
$-4\phantom{-}$ & $-6\phantom{-}$ &  $-10\phantom{-}$& 
{\it no}  &$\Psi=0$  \\
\hline
$\phantom{\Bigg|}$$\phantom{\Bigg|}$ & $1$ & $0$ & $1$ & 
$-3\phantom{-}$ & $-5\phantom{-}$ &  $-9\phantom{-}$ &
 {\it no}  &$\Psi=0$  \\
\hline
$\phantom{\Bigg|}$$\phantom{\Bigg|}$ & $0$ & $0$ & $1$ & 
$-10\phantom{-}$ & $-1\phantom{-}$ &  $-9\phantom{-}$ & 
 {\it no} &  $\Lam,\Psi=0$ \\
\hline
$\phantom{\Bigg|}$$\phantom{\Bigg|}$ & $0$ & $0$ & $1$ & 
$-8\phantom{-}$ & $-6\phantom{-}$ &  $-6\phantom{-}$&
 {\it no} &$\Psi=0$  \\
\hline
\end{tabular}$~$\\\emph{continued next page}
\begin{tabular}{|c|c|c|c|c|c|c|c|c|}
\hline
$\phantom{\Bigg|}${\bf{by}}$\phantom{\Bigg|}$&                     
         ${\bsym{x}}$                           &                  
         ${\bsym{y}}$                           &                  
         ${\bsym{z}}$                           &                  
         ${\bsym{X_{L^1}}}$                     &
         ${\bsym{X_{L^2}}}$                     &
         ${\bsym{X_{L^3}}}$                     & 
          {
           \begin{tabular}{c} 
             {\bf Compatible}         \\
             {\bf with Exp.} \end{tabular} }                   &
          {\bf O.K. if ...} \\
\hline
$\phantom{\Bigg|}$$\phantom{\Bigg|}$ & $1$ & $0$ & $1$ & 
$-8\phantom{-}$ & $-4\phantom{-}$ &  $-5\phantom{-}$& 
  {\it no} &$\Lam,\Psi=0$ \\ 
\hline
$\phantom{\Bigg|}$$\phantom{\Bigg|}$ & $1$ & $0$ & $1$ & 
$-6\phantom{-}$ & $-9\phantom{-}$ &  $-2\phantom{-}$&
 {\it no} & $\Psi=0$  \\
\hline
$\phantom{\Bigg|}$$\phantom{\Bigg|}$ & $2$ & $0$ & $1$ & 
$-8\phantom{-}$ & $-2\phantom{-}$ &  $-4\phantom{-}$& 
  {\it no} &$\Lam,\Psi=0$ \\
\hline
$\phantom{\Bigg|}$$\phantom{\Bigg|}$ & $1$ & $0$ & $1$ & 
$-10\phantom{-}$ & $-7\phantom{-}$ &  $-9\phantom{-}$&
 \phantom{$^{\ast}$}{\it no}$^{\ast}$ &$\Psi=0$  \\
\hline
$\phantom{\Bigg|}$$\phantom{\Bigg|}$ & $2$ & $0$ & $1$ & 
$-10\phantom{-}$ & $-5\phantom{-}$ &  $-8\phantom{-}$&
 \phantom{$^{\ast}$}{\it no}$^{\ast}$ & $\Psi=0$  \\
\hline
$\phantom{\Bigg|}$$\phantom{\Bigg|}$ & $2$ & $0$ & $1$ & 
$-8\phantom{-}$ & $-10\phantom{-}$ &  $-5\phantom{-}$&
 \phantom{$^{\ast}$}{\it no}$^{\ast}$ 
  &$\Psi=0$   \\
\hline
$\phantom{\Bigg|}$\cite{mnrv}$\phantom{\Bigg|}$ & $1$ & $0$ & 
$1$ & $-
\frac{113}{30}\phantom{-}$ & $-
\frac{113}{30}\phantom{-}$ &  $-
\frac{113}{30}\phantom{-}$ & {\it yes} & \\
\hline
$\phantom{\Bigg|}$\cite{r}$\phantom{\Bigg|}$ & $0$ & $0$ & $0$ & $-
\frac{19}{5}\phantom{-}$ 
& $-
\frac{19}{5}\phantom{-}$ &  $-
\frac{19}{5}\phantom{-} $ & {\it no}  &$\Lam,\Lam^\prime,\Psi=0$  \\
\hline
$\phantom{\Bigg|}$\cite{l1,blr}$\phantom{\Bigg|}$ & 0 & 0 & 0 &   
$X_{L^3}+3$   &  $X_{L^3}+1$  &  ...$\Big/-\frac{41}{15}$  & 
{\it yes}$\Big/${\it no}  & $\Lam,\Lam^\prime=0$ \\
\hline
$\phantom{\Bigg|}$ & 1 & 0 & 0 &   
$X_{L^3}+3 $    &  $X_{L^3}+1 $  &   ...$\Big/-\frac{239}{105}$  & 
{\it yes}$\Big/${\it no}  &$\Lam,\Lam^\prime=0$ \\
\hline
$\phantom{\Bigg|}$ & 2 & 0 & 0 &
$X_{L^3}+3 $    &  $X_{L^3}+1 $  &   ...$\Big/-\frac{11}{6}$  & 
{\it yes}$\Big/${\it no} & $\Lam,\Lam^\prime=0$ \\
\hline
$\phantom{\Bigg|}$ & 3 & 0 & 0 & 
$X_{L^3}+3 $    &  $X_{L^3}+1 $  &   ...$\Big/-\frac{7}{5}$  & 
{\it yes}$\Big/${\it no}
 & $\Lam,\Lam^\prime=0$ \\    
\hline
$\phantom{\Bigg|}$ & 0 & 0 & 1 & 
$X_{L^3}+3 $    &  $X_{L^3}+1 $  &   ...$\Big/-\frac{73}{35}$  & 
{\it yes}$\Big/${\it no} 
 & \\    
\hline
\end{tabular}$~$\\\emph{continued next page}
\begin{tabular}{|c|c|c|c|c|c|c|c|c|}
\hline
$\phantom{\Bigg|}${\bf{by}}$\phantom{\Bigg|}$&                     
         ${\bsym{x}}$                           &                  
         ${\bsym{y}}$                           &                  
         ${\bsym{z}}$                           &                  
         ${\bsym{X_{L^1}}}$                     &
         ${\bsym{X_{L^2}}}$                     &
         ${\bsym{X_{L^3}}}$                     & 
          {
           \begin{tabular}{c} 
             {\bf Compatible}         \\
             {\bf with Exp.} \end{tabular} }                   &
          {\bf O.K. if ...} \\
\hline
$\phantom{\Bigg|}$ & 1 & 0 & 1 & 
$X_{L^3}+3 $    &  $X_{L^3}+1 $  &   ...$\Big/-\frac{33}{20}$  & 
{\it yes}$\Big/${\it no} & \\   
\hline
$\phantom{\Bigg|}$ & 2 & 0 & 1 & 
$X_{L^3}+3 $    &  $X_{L^3}+1 $  &   ...$\Big/-\frac{11}{9}$  & 
{\it yes}$\Big/${\it no} 
 & \\   
\hline
$\phantom{\Bigg|}$ & 3 & 0 & 1 & 
$X_{L^3}+3 $    &  $X_{L^3}+1 $  &   ...$\Big/-\frac{4}{5}$  & 
{\it yes}$\Big/${\it no} & \\   
\hline
$\phantom{\Bigg|}$ & 0 & $-6\phantom{-}$ & 0 &   
$X_{L^3}+3 $    &  $X_{L^3}+1 $  &   ...$\Big/-\frac{59}{15}$  & 
{\it yes}$\Big/${\it no} & $\Lam,\Lam^\prime=0$ \\   
\hline
$\phantom{\Bigg|}$ & 1 & $-6\phantom{-}$ & 0 &   
$X_{L^3}+3 $    &  $X_{L^3}+1 $  &   ...$\Big/-\frac{347}{105}$  & 
{\it yes}$\Big/${\it no} & $\Lam,\Lam^\prime=0$ \\   
\hline
$\phantom{\Bigg|}$ & 2 & $-6\phantom{-}$ & 0 &  
$X_{L^3}+3 $    &  $X_{L^3}+1 $  &   ...$\Big/-\frac{41}{15}$  & 
{\it yes}$\Big/${\it no}
 &  $\Lam,\Lam^\prime=0$ \\    
\hline
$\phantom{\Bigg|}$ & 0 & $-6\phantom{-}$ & 1 &
$X_{L^3}+3 $    &  $X_{L^3}+1 $  &   ...$\Big/-\frac{109}{35}$  & 
{\it yes}$\Big/${\it no}
  & \\   
\hline
$\phantom{\Bigg|}$ & 1 & $-6\phantom{-}$ & 1 & 
$X_{L^3}+3 $    &  $X_{L^3}+1 $  &   ...$\Big/-\frac{51}{20}$  & 
{\it yes}$\Big/${\it no} &  \\
\hline
$\phantom{\Bigg|}$ & 2 & $-6\phantom{-}$ & 1 &  
$X_{L^3}+3 $    &  $X_{L^3}+1 $  &   ...$\Big/-\frac{91}{45}$  & 
{\it yes}$\Big/${\it no} &  \\
\hline
$\phantom{\Bigg|}$\cite{l2,blpr}$\phantom{\Bigg|}$ & 0& 0 & 1 & $5$ 
& $2$ 
&$-2\phantom{-}  $  & {\it yes}  & \\
\hline
$\phantom{\Bigg|}$ & 1 & 0 & 1 & $5$ & $2$ 
&$-2\phantom{-} 
  $  & {\it yes} &  \\
\hline
$\phantom{\Bigg|}$ & 2 & 0 & 1 & $5$ & $2$ 
&$-2\phantom{-}   
$  & {\it yes} &  \\
\hline
$\phantom{\Bigg|}$ & 3 & 0 & 1 & $5$ & $2$ 
&$-2\phantom{-}   
$  & {\it yes} & \\
\hline
\end{tabular}$~$\\\emph{continued next page}
\begin{tabular}{|c|c|c|c|c|c|c|c|c|}
\hline
$\phantom{\Bigg|}${\bf{by}}$\phantom{\Bigg|}$&                     
         ${\bsym{x}}$                           &                  
         ${\bsym{y}}$                           &                  
         ${\bsym{z}}$                           &                  
         ${\bsym{X_{L^1}}}$                     &
         ${\bsym{X_{L^2}}}$                     &
         ${\bsym{X_{L^3}}}$                     & 
          {
           \begin{tabular}{c} 
             {\bf Compatible}         \\
             {\bf with Exp.} \end{tabular} }                   &
          {\bf O.K. if ...} \\
\hline
$\phantom{\Bigg|}$ & 0 & $-6\phantom{-}$ & 1 & $5$ & 
$2$ &$-2\phantom{-}   $  & {\it yes}  & \\
\hline
$\phantom{\Bigg|}$ & 1 & $-6\phantom{-}$ & 1 & $5$ & $2$ &
$-2\phantom{-}   $  & {\it yes} &  \\
\hline
$\phantom{\Bigg|}$ & 2 & $-6\phantom{-}$ & 1 & $5$ & $2$ &
$-2\phantom{-}
   $  & {\it yes}  & \\
\hline
$\phantom{\Bigg|}$\cite{blpr}$\phantom{\Bigg|}$ & 0 & $0$ & 1 &
$\frac{9}{2}$ 
& $\frac{3}{2} 
$ &$-\frac{9}{2}\phantom{-}$  & {\it yes} & \\
\hline
$\phantom{\Bigg|}$ & 1 & $0$ & 1 &
$\frac{9}{2}$ 
& $\frac{3}{2} 
$ &$-\frac{9}{2}\phantom{-}$  & {\it yes} & \\
\hline
$\phantom{\Bigg|}$ & 2 &$0$ & 1 &
$\frac{9}{2}$ 
& $\frac{3}{2} 
$ &$-\frac{9}{2}\phantom{-}$  & {\it yes} & \\
\hline
$\phantom{\Bigg|}$ & 3 &$0$ & 1 &
$\frac{9}{2}$ 
& $\frac{3}{2} 
$ &$-\frac{9}{2}\phantom{-}$  & {\it yes} & \\
\hline
$\phantom{\Bigg|}$ & 0 &$-6\phantom{-}$ & 1 &
$\frac{9}{2}$ 
& $\frac{3}{2} 
$ &$-\frac{9}{2}\phantom{-}$  & {\it yes} & \\
\hline
$\phantom{\Bigg|}$ & 1 &$-6\phantom{-}$ & 1 &
$\frac{9}{2}$ 
& $\frac{3}{2} 
$ &$-\frac{9}{2}\phantom{-}$  & {\it yes} & \\
\hline
$\phantom{\Bigg|}$ & 2 &$-6\phantom{-}$ & 1 & 
$\frac{9}{2}$ 
& $\frac{3}{2} 
$ &$-\frac{9}{2}\phantom{-}$  & {\it yes} & \\
\hline
$\phantom{\Bigg|}$\cite{eir}$\phantom{\Bigg|}$ & 3 &0 & 0 & 
$\frac{79}{30}   
$ 
& $-\frac{11}{30}\phantom{-} $  
& $-\frac{11}{30}\phantom{-}$ & {\phantom{$_{h.o.}$}\it 
(no)}$_{h.o.}$  & 
$\Psi=0$  \\
\hline
 $\phantom{\Bigg|}$\cite{hm}$\phantom{\Bigg|}$ & 0&1 & 0 & 
0 & 0 & 0 &
 {\it no}  & \\
\hline
 $\phantom{\Bigg|}$\cite{bs}$\phantom{\Bigg|}$ & 0&1 & 0 & 
0 & 0 & 0 & 
{\it no} &  \\
\hline
\end{tabular}$~$\\\emph{continued next page}
\begin{tabular}{|c|c|c|c|c|c|c|c|c|}
\hline
$\phantom{\Bigg|}${\bf{by}}$\phantom{\Bigg|}$&                     
         ${\bsym{x}}$                           &                  
         ${\bsym{y}}$                           &                  
         ${\bsym{z}}$                           &                  
         ${\bsym{X_{L^1}}}$                     &
         ${\bsym{X_{L^2}}}$                     &
         ${\bsym{X_{L^3}}}$                     & 
          {
           \begin{tabular}{c} 
             {\bf Compatible}         \\
             {\bf with Exp.} \end{tabular} }                   &
          {\bf O.K. if ...} \\ \hline
$\phantom{\Bigg|}$\cite{ky}$\phantom{\Bigg|}$ & 0&0 & 1 & 
$X_{L^3}$  & 
$X_{L^3}$  &  ...  &  
{\it yes} &   \\
\hline
$\phantom{\Bigg|}$ & 1&0 & 1 &  $X_{L^3}$   
&$X_{L^3}$ &  ... &  {\it yes}  &  \\
\hline
$\phantom{\Bigg|}$ & 2&0 & 1 & $X_{L^3}$  
&  $X_{L^3}$ &   ...  &  {\it yes}  &  \\
\hline
 $\phantom{\Bigg|}$ & 3&0 & 1 &  $X_{L^3}$  
&  $X_{L^3}$ &   ...  &  {\it yes} &   \\
\hline
$\phantom{\Bigg|}$ & 0&0 & 1 &  $X_{L^3}+1$  
& $X_{L^3}$ &  ... &  {\it yes}   & \\
\hline
 $\phantom{\Bigg|}$ & 1&0 & 1 & $X_{L^3}+1$  
&$X_{L^3}$ &  ...  &  {\it yes}  &  \\
\hline
 $\phantom{\Bigg|}$ & 2&0 & 1 & $X_{L^3}+1$  
&$X_{L^3}$ &  ...  &  {\it yes}  &  \\
\hline
 $\phantom{\Bigg|}$ & 3&0 & 1 &  $X_{L^3}+1$ & $X_{L^3}$& ... &  
{\it yes}  
& \\
\hline
 \end{tabular}
\end{center}
$~$\\
{\bf Table A:} {\it In the table  above we present the catalogue of
Froggatt-Nielsen models of the type defined in Section~1,
parametrized by $x,y,z,X_{L^1},X_{L^2},X_{L^3}$. For the entries 
``\emph{yes}'', ``\emph{no}'', etc. and $\Psi=0$,  
$\Lambda^{\prime\prime}=0$
 etc.  see Section~6. Models where the ``\emph{no}''
is labeled by a $^\ast$ are  more in accord  with  the  bounds if  
$\tilde{m}>250$ GeV, one gets a  {\it (no)}$_{h.o.}$
instead.}

\begin{center}

\begin{tabular}{|c|c|c|c|c|c|c|}
\hline 
$\phantom{\Bigg|}\!${\bf{by}}$\!\phantom{\Big|}$               
      &
${\bsym{X_{H^\mcal{D}}}},{\bsym{X_{H^\mcal{U}}}}$  &
${\bsym{X_{10^i}}}$                                            &
${\bsym{X_{\ol{5^i}}}}$                                  &
${\bsym{X_{\ol{N^i}}}}$                                  &
$\bsym{\eps}$                                              &
\begin{tabular}{c} {\bf{ 
Compatible with}}\\ {\bf{ Constraints?}}
\end{tabular}
\\
\hline
$\phantom{\Big|}$[16]& $2q_3,-2q_3$ &$3+q_3,2+q_3,q_3$&
$1+d_3,d_3,d_3$   &$ -$ & $0.22$ &  {\it no} 
\\
\hline
$\phantom{\Big|}$\cite{buya}& $0,0$ &$2,1,0$&
$1,0,0$   &$ 0,1,2$ & $\frac{1}{\sqrt{300}}$ &  {\it no} 
\\
\hline
$\phantom{\Big|}$& $0,0$ &$2,1,0$&
$1,0,0$   &$ 0,0,2$ & $\frac{1}{\sqrt{300}}$ &   {\it no} \\
\hline
$\phantom{\Big|}$& $0,0$ &$2,1,0$&
$1,0,0$   &$ 1,1,2$ & $\frac{1}{\sqrt{300}}$ &   {\it no} \\
\hline
$\phantom{\Big|}$& $0,0$ &$2,1,0$&
$2,1,1$   &$ 0,0,1$ & $\frac{1}{\sqrt{300}}$ &   {\it no} \\
\hline
$\phantom{\Big|}$& $0,0$ &$2,1,0$&
$1,0,0$   &$0,0,1$ & $\frac{1}{\sqrt{300}}$ &   {\it no} \\
\hline
$\phantom{\Big|}$\cite{shafi}& $0,0$ &$3,2,0$&
$2,0,0$   & {\bf{--}} & $0.22$ &   {\it no} \\
\hline
$\phantom{\Big|}$& $0,0$ &$3,2,0$&
$3,1,1$   &{\bf{--}} & $0.22$ &   {\it no} \\
\hline
$\phantom{\Big|}$& $0,0$ &$3,2,0$&
$4,2,2$   & {\bf{--}} & $0.22$ &   {\it no} \\
\hline
$\phantom{\Big|}$\cite{tani}& $0,0$ &$3,2,0$&
$2,0,0$   & $2,0,0$ & $0.22$ &   {\it no} \\
\hline
$\phantom{\Big|}$\cite{sato}& $0,0$ &$2,1,0$&
$1,0,0$   & $2,1,0$ & $0.07$ &   {\it no} \\
\hline
$\phantom{\Big|}$& $0,0$ &$2,1,0$&
$2,1,1$   & $2,1,0$ & $0.07$ &   {\it no} \\
\hline
$\phantom{\Big|}$& $0,0$ &$2,1,0$&
$1,0,0$   & $0,0,0$ & $0.07$ &   {\it no} \\
\hline
$\phantom{\Big|}$& $0,0$ &$2,1,0$&
$2,1,1$   & $0,0,0$ & $0.07$ &   {\it no} \\
\hline
$\phantom{\Big|}$& $0,0$ &$3,1,0$&
$2,1,1$   & $2,1,0$ & $0.07$ &   {\it no} \\
\hline
$\phantom{\Big|}$& $0,0$ &$3,1,0$&
$2,1,1$   & $0,0,0$ & $0.07$ &   {\it no} \\
\hline
$\phantom{\Big|}$\cite{hitoshi}& $0,0$ &$2,1,0$&
$1,1,1$   & $0,0,0$ & $0.05$ &   {\it no}\\\hline
\end{tabular}
\end{center}
{\bf Table B:} {\it The models in this table do not lead to a
fermionic mass spectrum as in Eqs.~(\ref{1}-\ref{6}). 
Note that all models are $SU(5)$
invariant.}

\section{\label{illustration}Examining the Mass Matrices}
To demonstrate the validity of Eqs.~(\ref{oins}-\ref{fihr}), 
note that
the CKM matrix is nearly a unit-matrix, so that one should have
$\bsym{U^{(U_L)}}\approx\bsym{U^{(D_L)}}^\dagger$, {\it i.e.} these
two matrices should have ``the same Cabibbo structure'' (for a more
detailed analysis see Ref.~\cite{eir}). It is thus reasonable that
$\bsym{G^{(U)}}^\dagger{\bsym{G^{(U)}}}\approx\bsym{G^{(D)}}{\bsym
{G^{(D)}}}^\dagger$. This is indeed the case for the matrices
presented, {\it e.g.} Eq.~(\ref{droi}) gives
\begin{eqnarray}
\!\!\!\left(\begin{array}{ccc}
\eps^8 & \eps^6 & \eps^4 \\
\eps^6 & \eps^4 & \eps^2 \\
\eps^4 & \eps^2 & 1 \end{array}\right)^{\!\!T}\!\!\cdot\!\!\left(
\begin{array}{ccc}
\eps^8 & \eps^6 & \eps^4 \\
\eps^6 & \eps^4 & \eps^2 \\
\eps^4 & \eps^2 & 1 \end{array}\right)=~
\frac{1+\eps^4+\eps^8}{3}~\left(\begin{array}{lll}
\eps^4 & \eps^4 & \eps^4 \\
\eps^2 & \eps^2 & \eps^2 \\
1 & 1 & 1 \end{array}\right)\!\!\cdot\!\!\left(\begin{array}{lll}
\eps^4 & \eps^4 & \eps^4 \\
\eps^2 & \eps^2 & \eps^2 \\
1 & 1 & 1 \end{array}\right)^{\!\!T}\!\!.
\end{eqnarray}
Furthermore the $\bsym{G^{(...)}}$ have the correct  
eigenvalues\footnote{As the matrix $\bsym{G^{(...)}}\cdot
\bsym{G^{(...)}}^\dagger$  is Hermitian, it can be diagonalized 
with  its eigenvalues on the diagonal.  Thus the matrix  
${\bsym{D^{(...)}}}$ which one gets when bi-unitarily diagonalizing 
$\bsym{G^{(...)}}$  contains the square roots of  the  
eigenvalues of 
$\bsym{G^{(...)}}\cdot\bsym{G^{(...)}}^\dagger$ 
(times a phase) on its 
diagonal. This should not be confused
with the strictly speaking wrong statement that  
${\bsym{D^{(...)}}}$  
contains the eigenvalues of $\bsym{G^{(...)}}$. As a demonstration, 
consider the following matrix, its coefficients having been randomly 
generated with {\it Mathematica}$^\copyright$, their absolute values 
being in the interval $\Big[\frac{1}{\sqrt{10}},\sqrt{10}\Big]$:
\beqn
\bsym{\mathcal{M}}~=~
\left(\begin{array}{ccc}
 \phantom{-} 0.8614 - 2.9121~i  & -1.2754 + 0.1818~i &\phantom{-} 
0.6656 - 2.2503~i\\ 
 -1.2074 - 1.3980~i  & \phantom{-} 1.2850 - 0.3739~i &\phantom{-} 
0.5473 + 0.8524~i\\
 -0.2591 + 0.5662~i  & -0.4475 + 0.0637~i &-0.6963 + 1.5219~i
\end{array}\right)\nonumber
\eeqn
The absolute values of the eigenvalues of $\bsym{\mathcal{M}}$
are 2.6456, 2.5975, 1.7199. 
However, the  absolute values of the square roots of the 
eigenvalues of 
$\bsym{\mathcal{M}\mathcal{M}}^\dagger$ are  4.3262, 
2.4631, 1.1091. 
On the other hand, if the entries of $\mathcal{M}$ are 
$\epsilon$-suppressed, 
e.g. with the exponents
\beqn
\left(\begin{array}{ccc}    
5 & 3 & 2 \\
4 & 2 & 1 \\
3 & 1 & 0 \\ 
\end{array}\right),\nonumber  
\eeqn
one gets 2.3586, 0.0813, 0.0051  and  2.3701, 0.0819, 0.0050,
 respectively, 
almost the same.   } and hence reproduce the correct masses. For 
example, 
$\bsym{G^{(U)}}$ in Eq.~(\ref{oins}) has the  characteristic 
polynomial 
\begin{eqnarray}
-\det\bsym{{g^{(U)}}}~\eps^{12}~+~\Big[\big(
{g^{(U)}}_{\!22}~~{g^{(U)}}_{\!33}~-
~{g^{(U)}}_{\!23}~~{g^{(U)}}_{\!32}\big)
~~\eps^4~\nonum\\
\big({g^{(U)}}_{\!11}~~{g^{(U)}}_{\!33}~-~{g^{(U)}}_{\!13}~~
{g^{(U)}}_{\!31}\big)~~\eps^8~~+~~
\big({g^{(U)}}_{\!11}~~{g^{(U)}}_{\!22}~-~{g^{(U)}}_{\!12}~~
{g^{(U)}}_{\!21}\big)~~\eps^{12}~~\Big]~~\lam~&~&\nonum\\
-\big({g^{(U)}}_{\!11}~~\eps^8~+~{g^{(U)}}_{\!22}~~
\eps^4~+~~{g^{(U)}}_{\!33}\big)~~\lam^2~+~\lam^3&=&0.\nonum\\
\end{eqnarray}
Assuming that
\begin{eqnarray}
&&\det\bsym{{g^{(U)}}}\approx1\,,\nonum\\
&&{g^{(U)}}_{\!22}~~{g^{(U)}}_{\!33}~-~{g^{(U)}}_{\!23}~~
{g^{(U)}}_{\!32}
\approx1\,,\nonum\\
&&{g^{(U)}}_{\!33}~~{g^{(U)}}_{\!11}~-~{g^{(U)}}_{\!31}
~{g^{(U)}}_{\!13}~\approx1\,,\nonum\\
&&{g^{(U)}}_{\!11}~~{g^{(U)}}_{\!22}
-{g^{(U)}}_{\!12}~~{g^{(U)}}_{\!21}~\approx1\,,\nonum\\
&& {g^{(U)}}_{\!33}
\approx {g^{(U)}}_{\!22}~\approx {g^{(U)}}_{\!11}~\approx1\,,
\end{eqnarray}
this reduces to 
\begin{eqnarray}
\lam^3-(\eps^8+\eps^4+1)~\lam^2+(\eps^4+\eps^8+
\eps^{12})~\lam-\eps^{12}\approx0\,,
\end{eqnarray}
which can be factored as 
\begin{eqnarray} 
(\lam-1)(\lam-\eps^4)(\lam-\eps^8)\approx0\,,
\end{eqnarray}
thus reproducing the correct ratio of the eigenvalues, see
Eq.~(1.5). Alternatively, in the expression for the characteristic
polynomial one can also neglect higher order terms and only demand
\begin{eqnarray}
&&{g^{(U)}}_{\!33}~\approx1,\nonum\\ 
&&{g^{(U)}}_{\!22}~{g^{(U)}}_{\!33}~-~{g^{(U)}}_{\!23}~~
{g^{(U)}}_{\!32}
\approx 1,\nonum\\ 
&&\det\bsym{{g^{(U)}}}\approx 1,\end{eqnarray}
so that  \begin{eqnarray}  
\lam^3-\lam^2+\eps^4\lambda-\eps^{12}=0\,.
\end{eqnarray} 
Using $\eps=0.22$ we thus obtain $\lam_1=\eps^{8.00212},~\lam_2=\eps^
{3.99},~\lam_3=\eps^{0.001}$, which is also in good agreement.

To numerically demonstrate that $\bsym{G^{(U)}},\bsym{G^{(D)}}$ of
Eq.~(\ref{oins}) give the correct CKM-matrix, random $\mcal{O}\!(1)$
coefficients were generated with {\it Mathematica}$^\copyright$. 
With 
Eq.~(\ref{oins}) we obtained
\begin{eqnarray}\nonum
\bsym{{G^{(U)}}}&=&\left(\begin{array}{rrr} 
 1.41~\eps^8 & -0.95~\eps^5 & -1.73~\eps^3 \\
 0.94~\eps^7 & -1.19~\eps^4 & 1.12~\eps^2  \\
 1.35~\eps^5 & \phantom{-}2.10\phantom{~\eps^4}  & 1.52~\eps^2 
\end{array}\right),\nonum\\
\bsym{{G^{(D)}}}&=&
\left(\begin{array}{rrr} 
 1.22~\eps^4 & -1.45~\eps^3 & -1.76~\eps^3 \\
 0.69~\eps^3 & -0.94~\eps^2 & 0.60~\eps^2  \\
 -1.02~\eps & 0.68\phantom{~\eps^2}  & -1.97\phantom{~\eps^2} 
\end{array}\right).
\end{eqnarray}
Using $\eps=0.22$, this leads to  
\begin{equation}
m_b:m_s:m_d=-1.98:-0.68~\eps^2 :
1.13~\eps^4\,, 
\end{equation} 
\begin{equation}
m_t:m_c:m_u=-1.51:0.33~\eps^4 :2.56~\eps^8\,, 
\end{equation}
 and
\begin{eqnarray}\label{numckm}
\bsym{U^{C\!K\!M}}=\left(\begin{array}{rrr} 
 -0.98~~~ & -0.62~\eps~ & 3.66~\eps^3 \\
  0.62~\eps~ &  -0.98~~~ &   - 0.89~\eps^2\\
 4.18~\eps^3 & -0.77~\eps^2 & 0.99~~~ \end{array}\right),
\end{eqnarray}
which is the correct order of magnitude.

Finally, for completeness it should be mentioned that
Eqs.~(\ref{oins}) and (\ref{droi}) can also be generated from models
with a vector-like pair of Froggatt-Nielsen fields $A,B$, see
Ref.~\cite{dps}. A vector-like pair of Froggatt-Nielsen fields can
furthermore generate phenomenologically viable mass matrices which
cannot be generated from models with only one Froggatt-Nielsen field:
{\it e.g.} $X_{Q^1}=X_{\ol{U^1}}=-8$, $X_{Q^2}=X_{\ol{U^2}}=2$, 
$X_{Q^3
}=X_{\ol{U^3}}=X_{H^\mcal{U}}=0$ and $\big\langle{{A}}\big\rangle
=\big
\langle{{B}}\big\rangle$, see Ref.~\cite{ir}, \,
give\footnote{However
Froggatt-Nielsen models leading to such a matrix shall not be
considered in this paper as it cannot be achieved with only one
Froggatt-Nielsen-field $A$, because
\begin{eqnarray}
\left(\begin{array}{ccc} 
{16} &6  &8\\
{6}  &4  &2\\
{8}  &2  &0\end{array}\right)_{\!\!ij}=X_{Q^i}+X_{H^\mcal{U}}+
X_{\ol{U^j}}\nonum\end{eqnarray}
does not have a solution.}
\begin{eqnarray}
\bsym{{G^{(U)}}}\propto\left(\begin{array}{ccc} 
\eps^{16}  &\eps^{6}  &\eps^{8}\\
\eps^{6}  &\eps^4  &\eps^2\\
\eps^{8}  &\eps^2  &\eps^0\end{array}\right).
\end{eqnarray}
This leads to a characteristic polynomial of approximately the form
\begin{eqnarray}
\lam^3-\Big(1+\eps^4+\eps^{16}\Big)\lam^2+\Big(\eps^4
+\eps^{12}+\eps^{16}+\eps^{20}\Big)\lam-\Big(\eps^{12}
+\eps^{16}+\eps^{20}\Big)=0,\end{eqnarray}
with $\eps=0.22$ giving 
\begin{equation}
\lam_1=1,~~\lam_2=\eps^{4.00156},~~\lam_3=\eps^{7.99689}
\end{equation}
and thus 
\begin{equation}m_u:m_c:m_t=\eps^{8}:\eps^{4}:1,
\end{equation}
as required.

\end{appendix}
%
%

\end{document}